\documentclass[11pt,preprint]{aastex}

\def\gax{\mathrel{\raise.3ex\hbox{$>$}\mkern-14mu\lower0.6ex\hbox{$\sim$}}}
\def\lax{\mathrel{\raise.3ex\hbox{$<$}\mkern-14mu\lower0.6ex\hbox{$\sim$}}}
\def\gtorder{\mathrel{\raise.3ex\hbox{$>$}\mkern-14mu
             \lower0.6ex\hbox{$\sim$}}}
\def\ltorder{\mathrel{\raise.3ex\hbox{$<$}\mkern-14mu
             \lower0.6ex\hbox{$\sim$}}}
\def\he0435{HE~0435--1223}

\def\kbar{\langle\kappa\rangle}
\def\avgm{\langle M\rangle}

\def\bestab{\Delta t_{AB}=-8.00_{-0.82}^{+0.73}}
\def\bestac{\Delta t_{AC}=-2.10_{-0.71}^{+0.78}}
\def\bestad{\Delta t_{AD}=-14.37_{-0.85}^{+0.75}} 

\begin{document}

\title{The Time Delays of Gravitational Lens \he0435:\\ An Early-Type
Galaxy With a Rising Rotation Curve\footnote{Based on observations
obtained with the 1.3m telescope of the Small and Moderate Aperture
Research Telescope System (SMARTS), which is operated by the SMARTS
Consortium; the Apache Point Observatory 3.5m telescope, which is
owned and operated by the Astrophysical Research Consortium; and the
NASA/ESA {\it Hubble Space Telescope} as part of program HST-GO-9744
of the Space Telescope Science Institute, which is operated by the
Association of Universities for Research in Astronomy, Inc., under
NASA contract NAS~5-26555.}}

\author{C.S.\ Kochanek, N.D.\ Morgan}
 
\affil{\small Department of Astronomy, The Ohio State University, 140
West 18th Avenue, Columbus OH 43210}

\author{E.E.\ Falco, B.A.\ McLeod, J.N.\ Winn\footnote{Hubble Fellow}}

\affil{\small Harvard-Smithsonian Center for Astrophysics, 60 Garden
Street, Cambridge MA 02138}

\author{J. Dembicky, B.\ Ketzeback}

\affil{\small Apache Point Observatory, P.O. Box 59, Sunspot NM 88349}

\begin{abstract}
We present {\it Hubble Space Telescope} images and 2 years of optical
photometry of the quadruple quasar \he0435. The time delays between
the intrinsic quasar variations are $\bestad$, $\bestab$, and
$\bestac$~days. We also observed non-intrinsic variations of
$\sim$$0.1$~mag~yr$^{-1}$ that we attribute to microlensing. Instead
of the traditional approach of assuming a rotation curve for the lens
galaxy and then deriving the Hubble constant ($H_0$), we assume
$H_0=(72\pm7)$~km~s$^{-1}$~Mpc$^{-1}$ and derive constraints on the
rotation curve. On the scale over which the lensed images occur
($1\farcs2=5h^{-1}$~kpc~$\simeq 1.5R_e$), the lens galaxy must have a
rising rotation curve, and it cannot have a constant mass-to-light
ratio. These results add to the evidence that the structures of
early-type galaxies are heterogeneous.
\end{abstract}


\section{Introduction}
\label{sec:introduction}

Variability of the multiple images of a gravitationally lensed quasar
results from two distinct phenomena: intrinsic flux variations of the
background quasar, and microlensing by stars or other compact masses
in the foreground galaxy. Intrinsic variations are seen in different
quasar images at different times, owing to the different optical path
length and gravitational time delay associated with each image. Time
delays have been measured in about 10 systems, with varying levels of
accuracy and difficulty of interpretation, as recently reviewed by
Kochanek \& Schechter~\cite{Kochanek04} and
Kochanek~\cite{Kochanek05}. Microlensing, by contrast, is an extrinsic
phenomenon, arising from the granularity of the lensing mass
distribution. The granularity causes the image magnification to become
a very complex function of the source position (a ``caustic
pattern''). As the source moves with respect to the pattern,
uncorrelated variability is observed in the lensed
images. Microlensing has been observed in many quasar light curves,
most notably in the intensively monitored systems Q~2237+0305 and
Q~0957+564 (see the recent review by Wambsganss 2005).

Observations of the intrinsic and extrinsic variations have
traditionally been sought for entirely different reasons. Most the
effort in measuring time delays has been motivated by the prospect of
determining the Hubble constant ($H_0$) independently of local
distance indicators (Refsdal~1964). Microlensing is traditionally seen
as either noise in the cosmological measurement, or a means of
studying the quasar emitting region and the microlens mass
function. Given some recent developments in both observational
cosmology and gravitational lensing theory, we find it useful to
regard both time delays and microlensing variability as complementary
probes of the structure and composition of galaxy halos over a
particularly interesting range of galactocentric distances.

In the Cold Dark Matter (CDM) paradigm, galaxies have an inner region
that is predominantly composed of baryons, and an outer region that is
predominantly composed of dark matter particles. Most of the dark
matter is smoothly distributed, but a modest fraction ($\sim$1\%)
exists in clumps, which are sometimes referred to as CDM
``satellites'' or ``substructures'' (e.g.
Kauffmann~\cite{Kauffmann93}, Moore et al.~\cite{Moore99}, 
Klypin et al.~\cite{Klypin99}, Bode, Ostriker \& Turok~\cite{Bode01},
Zentner \& Bullock~\cite{Zentner03}).  For
a massive early-type galaxy, the transition region between baryon
dominance and dark-matter dominance is typically a few effective radii
($R_e$) from the center. Observations that are
sensitive to the mass distribution in this transition region have
resulted in confusing and apparently contradictory picture. There is
much evidence suggesting that galaxies have nearly flat rotation
curves (isothermal lens models) extending from the inner regions (see,
e.g.\ Gerhard et al.~2001; Rusin \& Kochanek~2003; Winn, Rusin, \&
Kochanek 2004, Treu \& Koopmans~2004) but there are some interesting
counter-examples (e.g.\ Romanowsky et al.~\cite{Romanowsky03}, Treu \&
Koopmans~\cite{Treu2002p6}).

The typical Einstein radius of a gravitational lens galaxy also
happens to be several effective radii, and hence the multiple images
of a background quasar tend to occur within the
baryon~$\rightarrow$~dark-matter transition region. Lensing
observations are thereby capable of providing constraints on the mass
distribution in that region. Furthermore, while traditional dynamical
observations are sensitive to the total enclosed mass within some
radius, time delays and microlensing depend upon the {\it local}
surface density and its degree of granularity, which are often of
greater interest.

First, consider time delays. Kochanek~(2002) showed that time delays
depend upon a combination of $H_0$ and the surface mass density
$\kbar=\langle\Sigma\rangle/\Sigma_c$ near the images, with $\Delta t
\propto (1-\kbar)/H_0$ to lowest order.\footnote{The dimensionless
surface density $\kappa$ is the surface density $\Sigma$ divided by
the critical surface density $\Sigma_c\equiv c^2 D_{\rm OS}/4\pi G
D_{\rm OL}D_{\rm LS}$, where $D_{\rm OL}$, $D_{\rm OS}$ and $D_{\rm
LS}$ are the angular diameter distances between the Observer, Lens and
Source.} Kochanek (2003) analyzed the 4 systems that have a simple
lens geometry and for which accurate time delays, astrometry and
photometry were available: PG~1115+080, SBS~1520+530, CLASS~B1600+434,
and HE~2149-2745. He found that the lens
galaxies in those systems have similar surface densities (with a
scatter in $\kappa$ of less than 0.07), but together they present a
problem for either CDM theory or the consensus value of $H_0$. If the
Hubble constant is $H_0 = (72\pm7)$~km~s$^{-1}$~Mpc$^{-1}$, as
suggested by Cepheid-based measurements (Freedman et al.~2001) and
analyses of microwave background fluctuations (Spergel et al.~2003),
then all four lens galaxies must have surface densities $\kbar \simeq
0.2$, which is close to what one would expect from a model with a
constant mass-to-light ratio ($M/L$). Only if $H_0\simeq
50$~km~s$^{-1}$~Mpc$^{-1}$ can they have $\kbar \simeq 0.5$, as one
would expect from a galaxy with a flat rotation curve. To make
progress we should (1) improve upon the accuracies of many of the
existing time delay measurements, some of which have uncertainties of
20\% or worse; (2) measure delays in more systems, to see whether the
result is a fluke (and, by extension, whether galaxies are
heterogeneous); and (3) measure time delays in systems for which
independent dynamical measurements are available; and (4) test for the
effects and possible biases that are expected to be caused by
variations in lens galaxy environments, by measuring delays for lenses
in groups or clusters. 

Next, consider microlensing. The character of the time variability in
a microlensing light curve depends upon the local surface density at
the position of the quasar image, and upon the fraction of that
surface density that is composed of stars ($\kappa_\star/\kappa$). A
statistical analysis of the instantaneous flux ratios of an ensemble
of lenses has shown that there is a clear difference between the
magnifications of images that are minima of the time-delay surface,
and the magnifications of saddle-point images. This provides evidence
that the stars represent no more than about 20\% of the total surface
density of the lens galaxies at the positions of the quasar images
(Schechter \& Wambsganss~2002, Kochanek \& Dalal~2004). This, in turn,
supports the standard isothermal models (which have considerable dark
matter near the lensed images) and argues against constant-$M/L$
models. It should be possible to go beyond the ensemble analysis and
analyze the light curves of individual systems in detail, now that there
is a method for analyzing quasar
microlensing light curves of arbitrary complexity
(Kochanek~2004). This algorithm can be used to derive estimates of all
the interesting physical variables, including $\kappa$, $\kappa_*$ and
the mean stellar mass $\avgm$. Unfortunately, with the exception of
Q~2237+030, the necessary data for such analyses is lacking.
 
With these motivations, we have undertaken a campaign to monitor a
large number of gravitational lenses using a network of optical
telescopes. Our aim is to obtain accurate multi-year light curves for
approximately 25 systems, with a a time sampling of 1-2 points per
week whenever a target is observable. We also rely on observations
with the {\it Hubble Space Telescope} ({\it HST}\,) to provide the
accurate photometry and astrometry that are necessary for lens
modeling.  This paper, which examines the particular lens \he0435, is
the first in what we hope will be a long series of new or improved
time delay measurements, microlensing detections, and constraints on
the structures of lens galaxies.

The quadruple-image quasar \he0435 was discovered by Wisotzki et
al.~(2002). The background quasar (``source'') has a redshift of
$z_s=1.689$. The redshift of the lens galaxy was recently measured by
Morgan et al.~(2004) to be $z_l=0.4541$. Evidence of microlensing at
optical wavelengths was presented previously by Wisotzki et
al.~(2003), based on integral-field spectrophotometry. In the
following section, we present new {\it HST}\, images as well as a
re-analysis of previously presented images. In \S~\ref{sec:monitoring}
we discuss the design of our lens monitoring campaign and
data-reduction pipeline, and present 2~years of photometry of
\he0435. We introduce a new method for analyzing gravitational lens
light curves, which is designed to separate the intrinsic variations
from the microlensing variations, and to determine the time delays
between all 4 quasar images. In \S~\ref{sec:models-macro} we present a
comprehensive study of the constraints on the mass distribution of the
lens galaxy that are provided by the combination of time delay
measurements and the {\it HST}\, data. 
 In the final section, we summarize our
conclusions and draw a comparison with the results of other time-delay
measurements.

Unless otherwise stated, we assume a flat cosmological model with
$\Omega_{\rm M} = 0.3$. Given the source and lens redshifts, the
conversion from angular to physical scales is
$1\farcs0=4.05h^{-1}$~kpc (with $H_0 = 100h$~km~s$^{-1}$~Mpc$^{-1}$),
the critical surface density is $\Sigma_c=5.18 \times 10^{10}$~$h^{-1}
M_\odot/$arcsec$^2$, and the relation between velocity dispersion and
Einstein radius $b$ for a singular isothermal sphere (SIS) is $\sigma
= 235\sqrt{b/1\farcs0}$~km~s$^{-1}$.

\section{{\it Hubble Space Telescope} Observations}
\label{sec:hst}

\begin{figure}
\epsscale{0.6}
\plotone{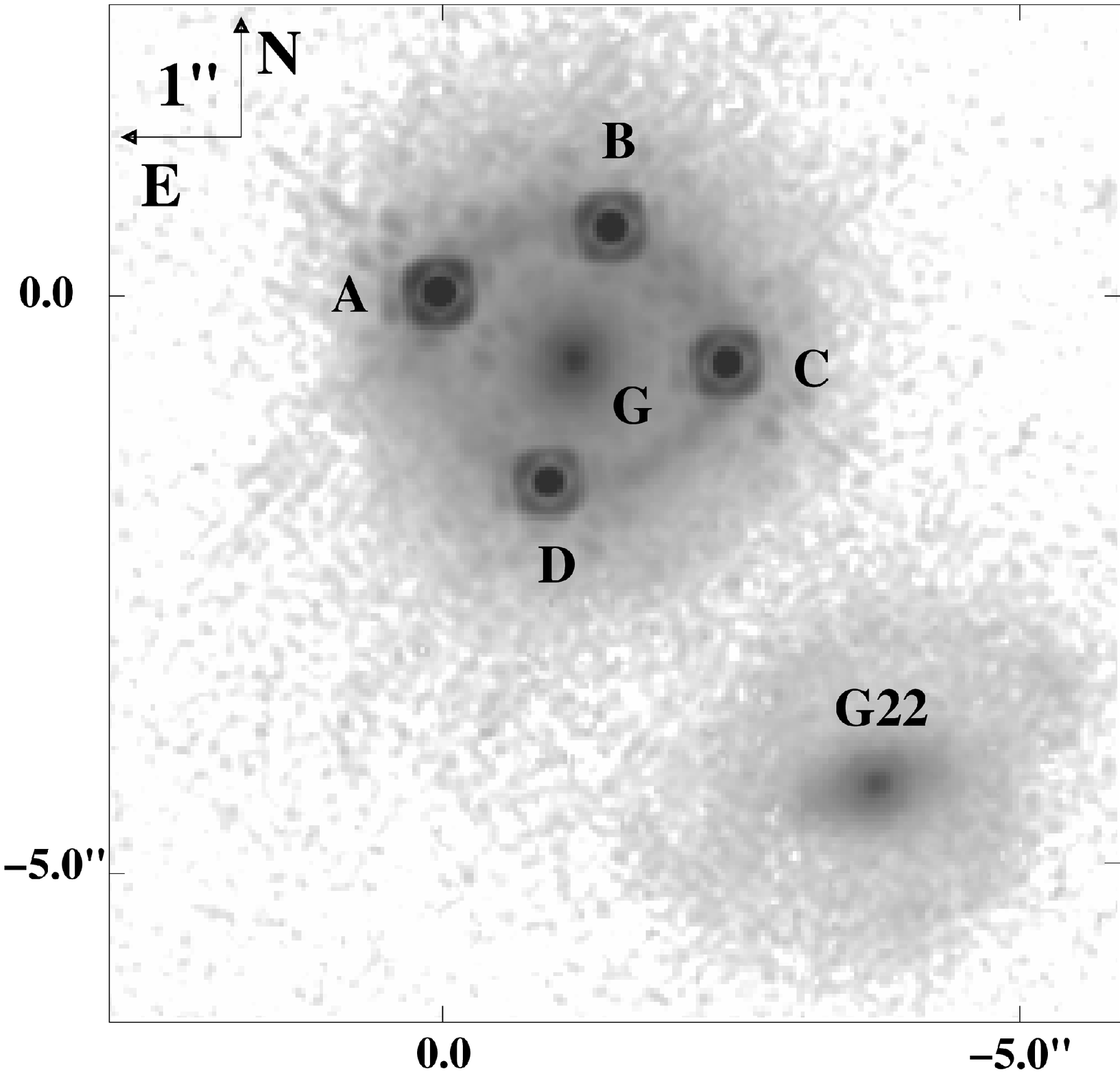}
\plotone{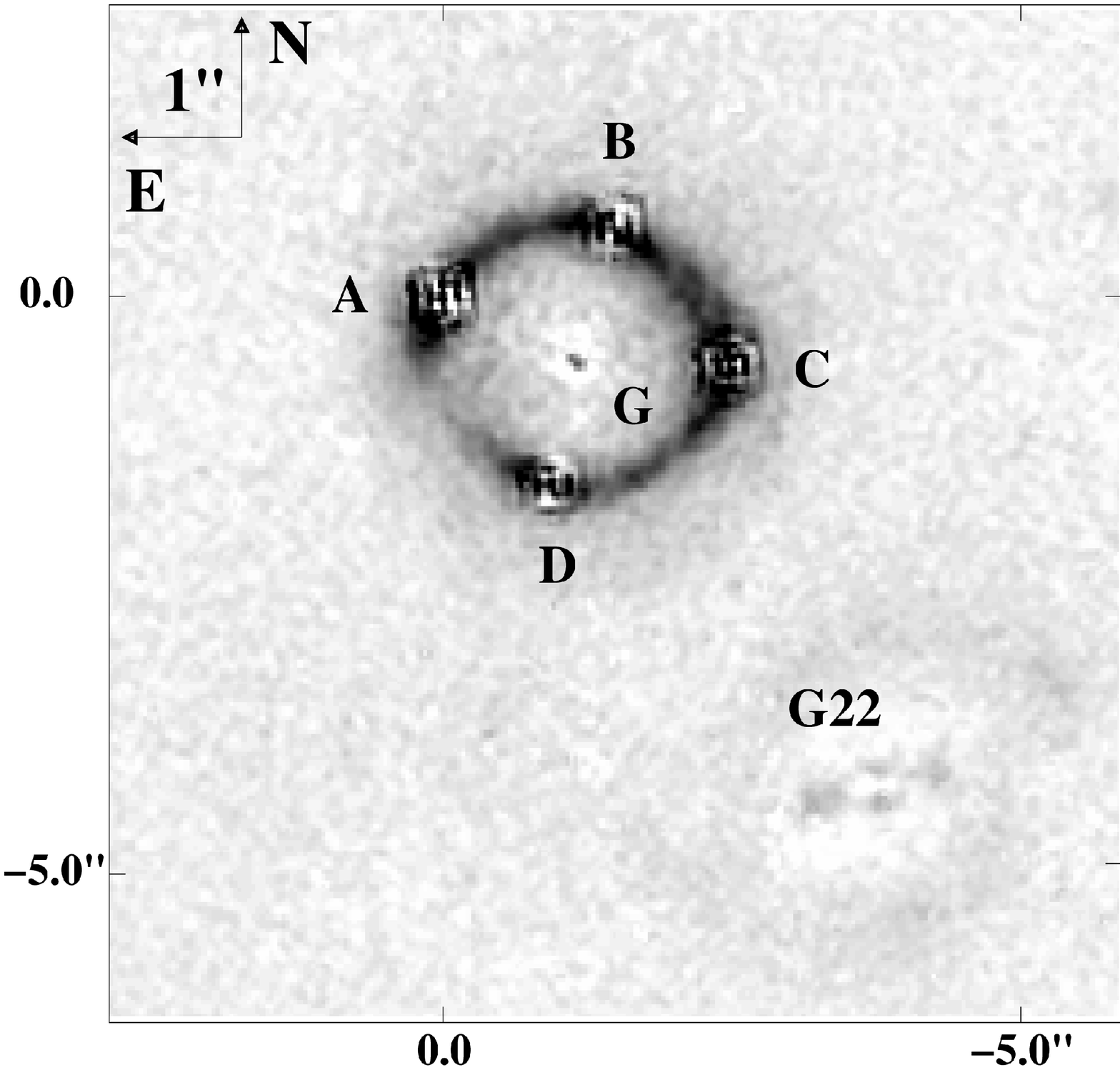}
\epsscale{1.0}
\caption{ The gravitational lens \he0435\ as observed with {\it
HST}/NICMOS in the H-band. The top panel shows the original image. The
bottom panel shows the residuals after subtracting a model of the 4
quasar images A--D, the lens galaxy G, and the neighboring galaxy
G22. Within about $0\farcs2$ of each quasar image position are
subtraction artifacts due to imperfections of the PSF model. }  
\label{fig:hband}
\end{figure}

We have observed \he0435 in the V (F555W), I (F814W) and H (F160W)
bands using {\it HST}. The 2000~s V-band and 1450~s I-band images were
both obtained as five dithered sub-images with the Wide Field Channel
(WFC) of the Advanced Camera for Surveys (ACS) on 2003~August~18. The
ACS images were reduced using the Pyraf-based MULTIDRIZZLE package.
The 2560~sec H-band image was obtained as four dithered sub-images on
2004~January~10 using the Near-Infrared Camera and Multi-Object
Spectrograph (NICMOS). These data were reduced using our own NICRED
software (see Leh\'ar et al.~2000).  Since the V and I band images
have already been presented by Morgan et al.~(2004), we focus here on
the new H-band image, shown in Fig.~\ref{fig:hband}.  We have labeled
the four quasar images A--D, the elliptical lens galaxy G, and a
nearby (SBb) spiral galaxy G22, following the nomenclature of Morgan
et al.\ (2004).  The quasar host galaxy has been stretched into a
nearly complete Einstein ring, which is prominent in the H-band image
after the quasars and lens galaxy have been subtracted.

We fitted the H-band image with a photometric model consisting of
point sources (representing the quasars), a de~Vaucouleurs bulge (the
lens galaxy), a de~Vaucouleurs bulge and an exponential disk (the
neighboring galaxy G22), and a lensed exponential disk for the host
galaxy.  All of these were
convolved with a PSF model and then a least-squares fit to the image
was performed, following the procedures of Leh\'ar et al. (2000).  The
PSF model was selected from a series of 8 images of bright stars that
were observed for this purpose (Yoo et al.\ 2005). We tried each of
the 8 stars as a PSF model and for the final analysis we selected the
star that resulted in the smallest residuals when applied to the
\he0435 image. In our previous experience with subtracting quasar
images with NICMOS, we have found that there are often significant
residuals near the Airy ring of the diffraction pattern. For this
reason, we included extra model parameters that allow for a distortion
of the PSF model, which resulted in a modest improvement. After
finding the best fit to the H-band image, we fitted the same model to
the V and I-band images, holding the astrometric and structural
parameters fixed and optimizing only the fluxes. In this manner we
estimated the colors of all the objects.  Table~\ref{tab:photom}
presents the astrometric and photometric results from an analysis of
the {\it HST}\, images.

The colors of the lens galaxy are in agreement with the predictions of
standard population synthesis models for early-type galaxies in which
star formation occurred at $z>1$. Using a range of such models to
model the full spectral energy distribution of the lens galaxy and its
evolution, we estimate that the rest-frame B-band absolute magnitude
of the galaxy is $M_B=-20.46\pm0.13$~mag (using
$H_0=100$~km~s$^{-1}$~Mpc$^{-1}$) 
and that it will evolve to an absolute B-band
magnitude of $M_B=-20.3\pm0.2$~mag at redshift zero. The neighboring
galaxy G22 is bluer than the main lens galaxy, and its colors are
better described by a model with a younger stellar population. Its
rest-frame B-band luminosity is $M_B=-20.5\pm0.5$ at the lens
redshift, evolving to $M_B=-20.4\pm0.5$ at redshift zero.  The
larger uncertainties than for the main lens galaxy are due to the 
broader range of evolution models consistent with the colors.
In the
model for the H-band image, the disk and the bulge of G22 have
comparable fluxes of H$\simeq 18.8$ and $18.7$ respectively
and scale lengths of $R_d \simeq 0\farcs67 \simeq 2.7h^{-1}$~kpc and
$R_e  \simeq 0\farcs37 \simeq 1.5h^{-1}$~kpc respectively.

The four quasar images have virtually identical colors, with the
exception that image C seems to be redder by $0.17\pm0.08$~mag in
V$-$H than the other images.  The colors of the quasar host galaxy
(see ``H'' in Table~\ref{tab:photom}) are consistent with those
expected of an actively star-forming galaxy or from a galaxy that
experienced a starburst at $z\simeq 2$.  Using the method
described by Kochanek, Keeton \& McLeod~(2000), we determined the
closed curve that tracks the peak brightness of the Einstein ring as
the azimuthal angle is varied from 0 to 2$\pi$ around the main lens
galaxy.  This curve is used as a modeling constraint in
\S~\ref{sec:models-macro}.

\section{Optical Monitoring}
\label{sec:monitoring}

\subsection{Observations and Data Reduction}
\label{subsec:monitoring-obs}

The photometric monitoring observations took place between December
2003 and September 2005. Almost all of the data were obtained with the
dual-beam ANDICAM camera (Depoy et al.~2003) mounted on the SMARTS
1.3m telescope, which is located at the Cerro Tololo Inter-American
Observatory, in Chile. On each night, we obtained three 5~minute
R-band exposures. We obtained simultaneous data in the J-band with
ANDICAM, but we do not present any analysis of those data because of
their much lower signal-to-noise ratio. A few observations were made
with SPICAM on the 3.5m telescope of Apache Point Observatory (APO),
in New Mexico. These APO observations consisted of three 1.5~min
exposures in the Sloan Digital Sky Survey (SDSS) r-band at each
epoch. For our final analysis, we retained only those images with a
seeing of $1\farcs8$ or better.

Although the data for this particular target were obtained with only
two different telescopes (and are dominated by data from only one
telescope), our monitoring campaign generally relies upon a broad
array of different telescopes. For this reason, our data reduction
pipeline was designed to cope with very heterogeneous images. The
pixel scale, rotation, and other properties specific to the camera and
telescope are stored in the image headers and accessed by the
reduction pipeline. The basic idea underlying the pipeline is a
version of PSF fitting. We perform a least-squares fit to each image,
using a model with parameters that represent the quasar images, the
lens galaxy, a set of comparison stars, the sky level, and the point
spread function (PSF). The details are as follows.

For each target field, we define a set of subfields. One subfield
encompasses the gravitational lens, and each of the others is centered
on a comparison star. For the lens subfield, the model includes point
sources for the template stars and quasar images, and an approximate
de~Vaucouleurs profile (see below) for the lens galaxy, all of which
are convolved with the PSF.  For HE0435--1223, the nearby galaxy
G22 lay outside the modeled region.  We hold fixed the relative
positions of the quasar images and the structural parameters of the
lens galaxy (i.e.\ its effective radius, axis ratio, and position
angle), based on the parameters derived from {\it HST}\, images. For
the comparison star subfields, the model is a point source convolved
with the PSF model. In addition, each subfield is given an independent
sky level.

The PSF model is a superposition of three elliptical Gaussian
functions. The relative major axis widths of the three functions are held
constant, but the ellipticity and orientation of each function is
allowed to vary.  We allow for spatial gradients in each of the PSF parameters. 
We flux calibrate the images by including a prior constraint on the
fluxes of the comparison stars.  The mean flux of the comparison stars 
will vary slightly between frames (because of differences between the
flux ratios of the comparison stars in the prior and in the best fit
to the data).  These variations provide a simple means of checking
for significant problems in the PSF models.
In the rare cases when a comparison star
proves to be significantly variable, then either the data from that
star are discarded, or, if they are desired for the determination of
PSF parameters, they are kept and assigned very large flux
uncertainties. Subfields that contain saturated pixels or that lie too
close to the edge of the chip are ignored.

The lens galaxies are modeled using a Gaussian approximation to a
de~Vaucouleurs profile, i.e., a combination of a small number of
elliptical Gaussian functions that best fits the integrated light
profile of the appropriate de~Vaucouleurs function. The accuracy of
the approximation is controlled by the number of Gaussian functions
employed. In practice, a single Gaussian function is often sufficient,
since the lens galaxy typically contributes only a small fraction of
the total light of a lens system. The advantage of the Gaussian
decomposition is that the convolution with the PSF model can be
computed analytically, allowing for extremely rapid computation. The
integrated flux of the lens galaxy is required to be constant across
all epochs, with a value that provides the minimum $\chi^2$ when
applied to the entire series of images. 
A single Gaussian component was sufficient for HE0435--1223.

\begin{figure}[p]
\plotone{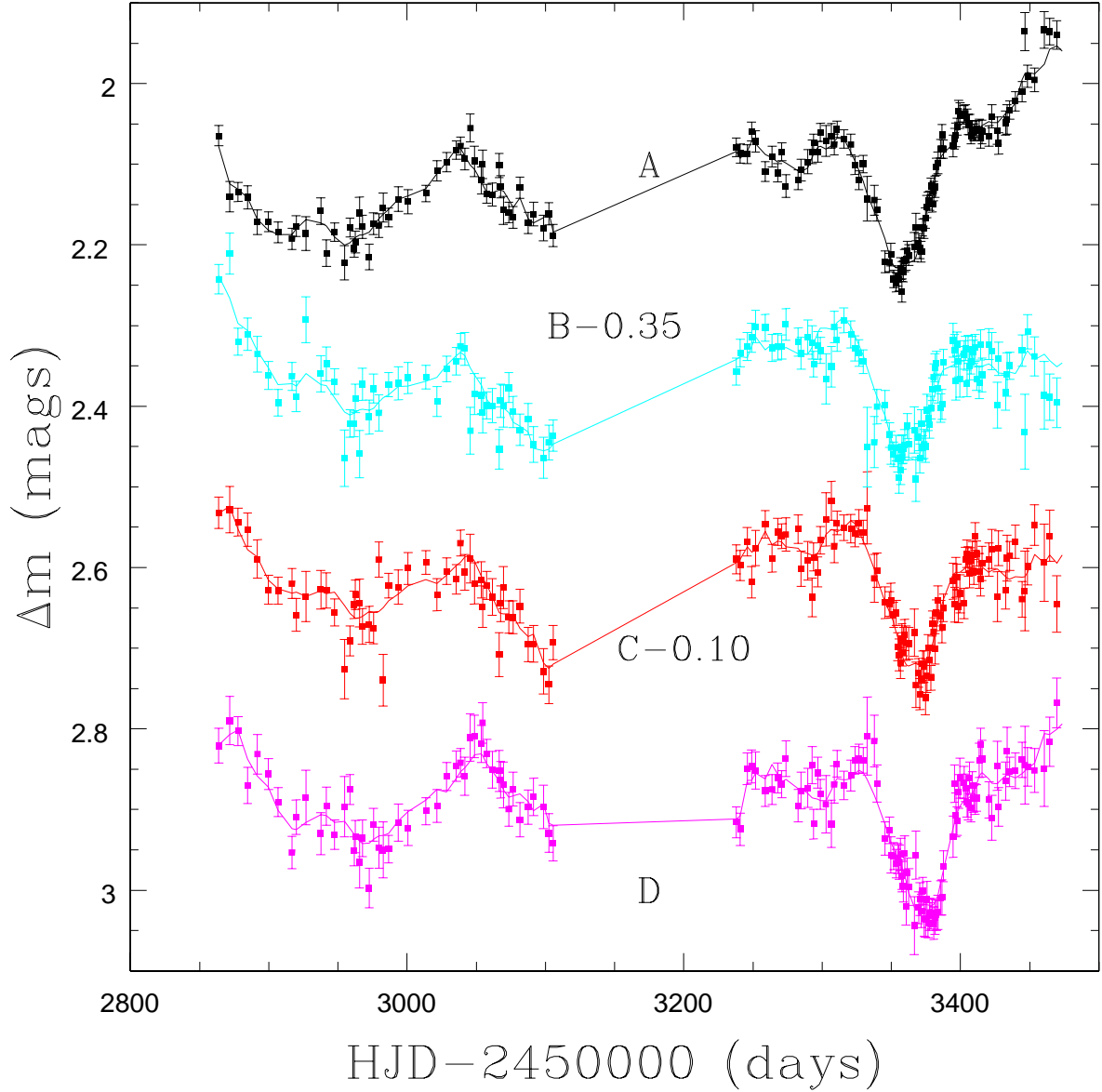}
\caption{The light curves of images A, B, C, and D of \he0435.
Arbitrary magnitude offsets have been applied to the B and C light
curves, for display purposes. The solid lines are the best-fitting
model light curves. }
\label{fig:lightcurves1}
\end{figure}

For \he0435, a listing of the comparison stars and their properties is
given in Table~\ref{tab:stars}. The photometry for the four quasar
images (A, B, C, and D) is given in Table~\ref{tab:lightcurves}, along
with the differential variability of the flux standards
and the value of $\chi^2/N_{dof}$, as
a measure of image quality and the success of the fitting procedure.
The r-band data were adjusted to the R-band scale by assuming an
offset of $0.032$~mag, which was estimated by comparing images in each
filter that were taken at similar epochs. The quoted uncertainties in
the best-fitting quasar fluxes were determined from the full
covariance matrix, and therefore incorporate the uncertainty the PSF
model (but not the uncertainty in the lens galaxy flux, which was held
constant after the optimization described in the previous
paragraph). In the subsequent analysis, the uncertainty in each point
with $\chi^2/N_{dof}>1$
was enlarged until the value of $\chi^2/N_{dof}$ for that particular
image was unity, in order to lower the statistical weight of the
points that were derived from problematic images.  The final
light curves are plotted in Fig.~\ref{fig:lightcurves1}.

\begin{figure}[p]
\plotone{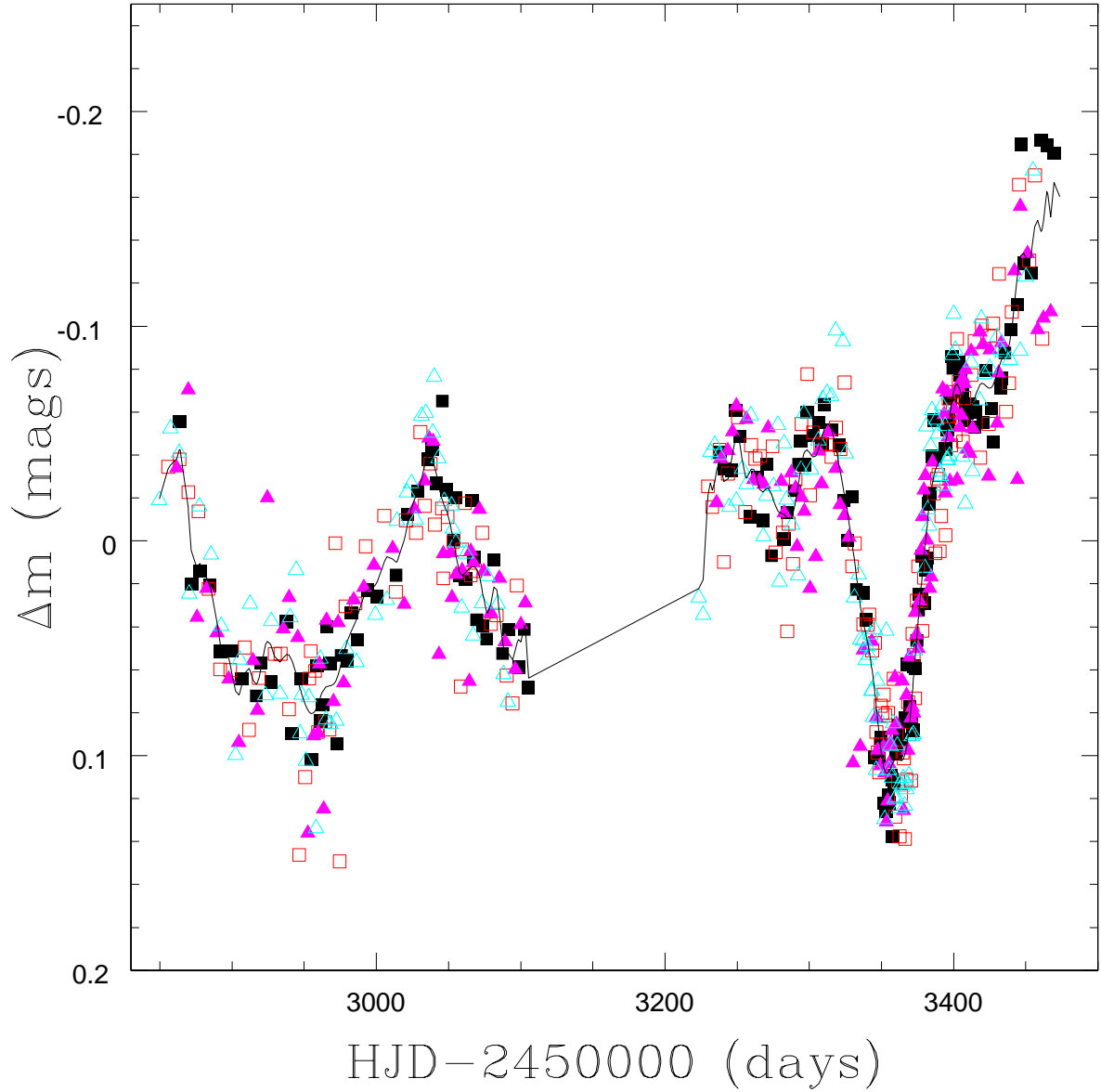}
\caption{The time-shifted light curves of images A, B, C, and D of
\he0435. The differential microlensing variations have been removed
from the light curves of images B, C, and D. The solid line is the
best-fitting model of the intrinsic variations of the source
quasar. Error bars are not shown, to avoid clutter. }
\label{fig:lightcurves2}
\end{figure}

\begin{figure}[p]
\plotone{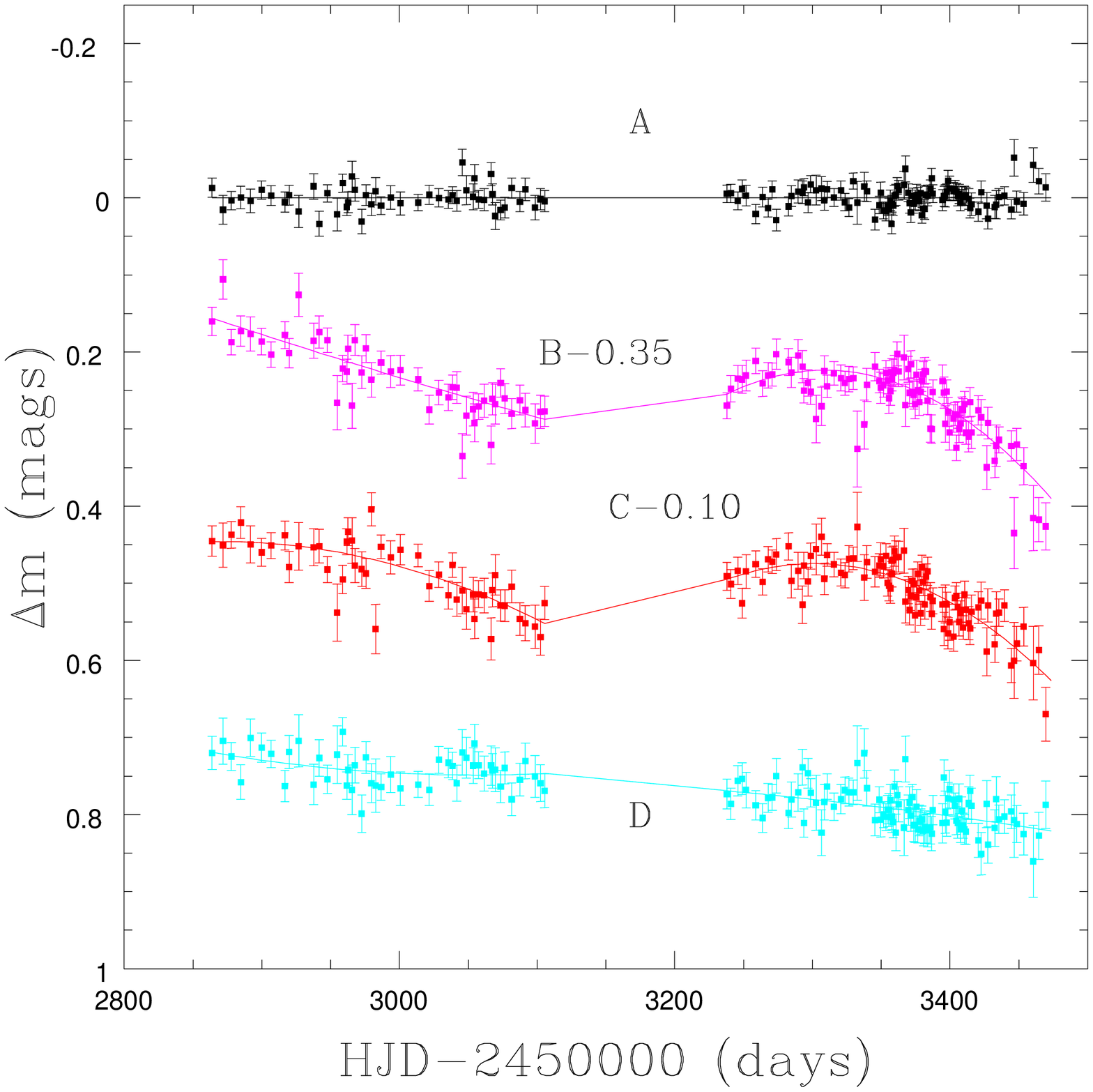}
\caption{The differential microlensing light curves of images A, B, C,
and D of \he0435, relative to image A. The best-fitting model of the
intrinsic variations of the source quasar has been subtracted from the
observed light curve of each image. }
\label{fig:lightcurves3}
\end{figure}

\subsection{Light Curve Analysis}
\label{subsec:lightcurves}

The patterns in the light curves that are common to all the images
indicate that we have observed intrinsic variability at the level of
0.2~mag, which is highly significant when compared to our typical
uncertainties of 0.01--0.02~mag.  Less obviously, there are
smaller-amplitude patterns that are specific to the light curve of
each image. These uncorrelated patterns are the hallmarks of
microlensing.  Our goal was to decompose each light curve into
intrinsic and the extrinsic variations, in order to estimate the
differential time delays between the images and to analyze the
microlensing variability.

We began with the premise that the intrinsic variations of the source
quasar can be approximated as a Legendre series,
\begin{equation}
     s(t) \simeq \sum_{m=0}^{N_{src}} a_m P_m \left[ { t - t_c \over \delta t }\right],
\end{equation}
where $s(t)$ is the magnitude of the source at time $t$, $t_c =
(t_N+t_1)/2$ is the midpoint of the time series, $\delta t =
(t_N-t_1)/2$ is the half-width of the time series, and $P_m$ is the
$m$th Legendre polynomial. In the absence of microlensing, the light
curve of the $i$th lensed image, $m_i(t)$, would be a time-shifted,
magnified copy of the source light curve.  To model the real light
curves, we adopt image A as a reference image, and assume that the
microlensing variations of each image relative to A can also be
written as a Legendre series in time. Thus, $m_i(t) = s(t + \Delta
t_i) + \Delta \mu_i(t)$, where we have defined
\begin{equation}
  \Delta\mu_i(t) = \sum_{m=0}^{N_\mu} c_{mi} P_m \left[ { t - t_c \over \delta t }\right],
\end{equation}
incorporating both the static magnification of the image and the
differential microlensing variations.  In fitting the observations
$m_{ij}$ of the magnitude of image $i$ at time $t_j$, the ordinary
fitting statistic is
\begin{equation}
   \chi^2 = \sum_{i=1}^{N_{im}} \sum_{j=1}^{N_{obs}} \left(
     {  m_{ij} - s(t_{j}+\Delta t_i) - \Delta\mu_i(t_{j}) \over \sigma_{ij} }\right)^2.
\end{equation}

We cannot simply proceed by minimizing $\chi^2$, because any choice of
the time delays would provide an acceptable fit if series of
arbitrarily high orders were allowed. Some kind of restriction must be
imposed that forces the model to exhibit ``reasonable'' quasar
variability. Our approach, which is similar to that of Press, Rybicki
\& Hewitt~\cite{Press1992p416}, is to apply an {\it a priori}
constraint that the intrinsic light curve has a power spectrum that is
approximately the power spectrum of a ``typical'' quasar.  Vanden~Berk
et al.~(2004) have measured the typical quasar structure function in the
r-band, using a large ensemble of quasar photometry from the SDSS,
finding
\begin{equation}
V(\tau)=(\tau/\tau_0)^\gamma \vspace{0.1in} {\rm mag}^2
\end{equation}
with $\tau_0\simeq$~70,000~days and $\gamma \simeq 1/3$. The $m$th
term in our Legendre series has a mean square magnitude variation of
$a_m^2/(2m-1)$ and a characteristic rest-frame time scale of
$t_m=4\delta t/m(1+z_s)$.  Our ``reasonability'' constraint is that
the mean squared power of each term should be
$(t_m/\tau_0)^{1+\gamma}$ (i.e.\ the root-mean-squared variations
should vary as $m^{-4/3}$ for $\gamma =1/3$.  Thus, instead of minimizing
$\chi^2$, we minimize
\begin{equation}
    H = \chi^2 + \lambda \sum_{m=0}^{N_{src}} { a_m^2 \over 2m-1}
        \left( {  \tau_0 \over t_m} \right)^{1+\gamma},
\end{equation}
where $\lambda$ is a Lagrangian multiplier that controls the weight of
this {\it a priori} constraint on the quasar structure function
relative to the weight of the individual data points.  We determine
the optimal values of $\Delta t_i$ by differentiating $H$ with respect
to the parameters and solving the resulting linear equations. The data
from each season are considered separately.

The task remains to choose the orders $N_{src}$ and $N_\mu$ of the
Legendre series, and the strength of the Lagrange multiplier
$\lambda$.  In order to evaluate whether an increase in $N_{src}$ or
$N_\mu$ is justified by the data, we use the F-test.  We find that
increasing $N_{src}$ results in a significant improvement until
$N_{src} = 20$ and thereafter ceases to be significant.  Most of the
improvement occurs in the range $5 \leq N_{src} \leq 10$.  As for the
microlensing series, the improvement is significant as $N_\mu$ is
increased to 3 and is not significant for larger choices.  Most of the
improvement occurs in the passage from $N_\mu=1$ (no microlensing) to
$N_\mu=2$ (trends that are linear in time). We verified that these
results do not depend strongly on the choice of $\lambda$, by
comparing the cases $\lambda=0.01$ and $\lambda=1$.  

In our final analysis, we adopted the values $N_{src}=60$, $N_\mu=3$,
and $\lambda=1$. The choice of such a large value of $N_{src}$ should
result in conservative uncertainties in the time delays.  With these
values, the longest delay is $\bestad$ days, the intermediate delay is
$\bestab$~days and the shortest delay is $\bestac$~days.  The
fractional uncertainties in the measurements are approximately 6\%,
10\%, and 35\%, respectively. Fig.~\ref{fig:lightcurves2} shows the
superposed light curves after shifting each light curve by the
best-fitting time delay and subtracting the model of the differential
microlensing variability. Table~\ref{tab:micro} gives the best-fitting
parameters of the microlensing models for each season, and
Fig.~\ref{fig:lightcurves3} shows the differential microlensing light
curves. Relative to image A, image D has steadily faded. Meanwhile,
images B and C faded more rapidly in the first season, then brightened
between the two seasons, reached a peak during the second season and
faded rapidly towards the end of the season. The common behavior of
images B and C strongly suggests that it is actually image A that
exhibits the greatest microlensing variability, although only the
differential effects are observable.

As we would expect from an analysis of light curves that are short compared to a typical Einstein
crossing time and show only low levels of variability, we cannot learn a 
great deal from the microlensing data as yet.  We analyzed the microlensing 
light curves using the Monte Carlo method of Kochanek~(2004) under 
the assumption that the average mass of the microlenses is within 
the range $0.2h^2 M_\odot \leq \langle M_*\rangle \leq 2h^2 M_\odot$. 
With this assumption, we do obtain a preliminary estimate for the physical size of
the accretion disk: $0.8 \ltorder r_{s15} \ltorder 5.4$ where
$r_{s15}=r_s/(10^{15} h^{-1}\hbox{cm})$.  If we assume that the
viscous energy release is radiated locally, then we can also 
estimate the ratio between the disk length scale $r_s$ and the gravitational 
radius $r_g=GM/c^2$ of the black hole to find that
$r_s/r_g \simeq 39 (L/L_E)^{1/2} r_{s15}^{-1/2})$ for a black hole
radiating with Eddington ratio $L/L_E$.   Thus, our estimate of the
disk scale length is a reasonable match to the hot, inner regions 
of an accretion disk.  The data do not yet justify a
more detailed microlensing analysis, but the prediction of all the
microlensing models is that the future microlensing variability should
be more dramatic than observed to date.

\section{Models and Interpretation}
\label{sec:models-macro}

\subsection{Analytic theory}
\label{subsec:analytic}

Kochanek~(2002) presented an analytic theory of time delays that
allows one to establish the quantitative connection between $H_0$ and
the surface density of the lens galaxy, without any detailed lens
modeling. It is based on a multipole-Taylor expansion of the lens
potential (see also Trotter, Winn, \& Hewitt 2000).  In this section,
we apply this theory to \he0435\ in order to draw a few robust
conclusions about the mass distribution of the lens galaxy.

For any pair of quasar images, we define $\langle\kappa\rangle$ as the
mean surface density of the lens galaxy within an annulus whose inner
and outer radii are the locations of the quasar images. It is often
convenient to refer to this quantity as the ``annular'' surface
density. We further define $\eta$ as the logarithmic slope of the
surface density within the annulus (i.e.\ $\kappa\propto
R^{\eta-1}$). We emphasize that both $\langle\kappa\rangle$ and $\eta$
are fundamentally {\it local} quantities, by which we mean that they
describe the mass distribution over the very small range of radii that
is spanned by the quasar images. Considering the pair of images for
which the time delay is largest (and hence the fractional uncertainty
in the time delay is smallest), and applying the procedure of
Kochanek~(2002), we find
\begin{equation}
   H_0 = (193\pm 25) (1-\langle\kappa\rangle) - (23\pm 3)\langle\kappa\rangle (\eta-1)
   \hspace{0.05in} \hbox{km s}^{-1}~\hbox{Mpc}^{-1}.
\end{equation}
This expression enforces a relation between $\eta$ and
$\langle\kappa\rangle$, for a given value of $H_0$.  For example,
assuming that $H_0=(72\pm7)$~km~s$^{-1}$~Mpc$^{-1}$, and that the mass
distribution of the lens galaxy is isothermal within the annulus
($\eta=2$), we may conclude that $\langle\kappa\rangle\simeq 0.56\pm
0.06$. If instead $\eta=3$ (a steeper mass distribution), then the
surface density must be somewhat smaller: $0.51\pm0.06$. Likewise, for
the case $\eta=0$ (a shallower mass distribution), the surface density
must be somewhat larger: $0.63\pm0.06$.

Through this analysis, we see immediately that \he0435\ is unlike most
of the other simple, isolated time-delay lenses.  Analyses of those
other systems have shown that the lens galaxies are compatible with
$H_0=(72\pm7)$~km~s$^{-1}$~Mpc$^{-1}$ only if their annular surface
densities are considerably smaller than 0.5.  For example, the lens
galaxy of PG~1115+080 must have $\langle\kappa\rangle\simeq 0.24\pm
0.11$ near its Einstein radius.  In contrast, the surface density of
\he0435\ at its Einstein radius is apparently larger than 0.5. An
alternate way to express this result is that the lens galaxy of
\he0435\ has a slightly rising rotation curve, whereas the other lens
galaxies have falling rotation curves.

Kochanek~(2002) also showed how to use the astrometry and the time
delay ratios to analyze the angular structure of the lens potential,
and in particular to determine the balance between internal and
external sources of shear.  The ``internal quadrupole fraction''
$f_{\rm int}$ is defined such that a pure external shear has $f_{\rm
int} = 0$, an isothermal ellipsoid has $f_{\rm int} = 0.25$, and an
ellipsoidal mass distribution that is contained completely within its
Einstein ring has $f_{\rm int} = 1$.  In this case we find that the
angular structure is dominated by the external shear. As estimated
from the astrometry, $f_{\rm int} = 0.14\pm0.04$.  This agrees well
with the estimate based on the time delay ratios, which is $f_{\rm
int} = 0.18 \pm 0.04$.

\subsection{Parametric Lens Modeling}
\label{subsec:parametric}

In this section and the few sections to follow, we present a suite of
calculations using traditional parametric lens models, in order to
answer more specific questions about the mass distribution of the lens
galaxy, the neighboring galaxies, and the possible presence of a group
halo that envelops the galaxies.  Whenever the results of these models
can be compared to the analytic theory presented in the previous
sections, they agree within about 10\%.

Our goal is to identify models that successfully describe the
positions of the quasar images relative to the lens galaxy, the
Einstein ring curve formed by the quasar host galaxy, and the measured
time delays.  As a very conservative estimate of the uncertainties in
the relative time delays, we doubled the statistical uncertainties
that were presented in \S~\ref{subsec:lightcurves}.  We fit the 
positions of the quasar images, the lens galaxy and the Einstein
ring curve using {\it lensmodel} (Keeton~\cite{Keeton2001}).

We used a three-component model, in which the components are the
primary lens galaxy G, the nearby galaxy G22, and an independent
external shear. We used a weak {\it a priori} constraint on the axis
ratio of the G model ($q=0.74\pm0.10$), which is intended to match the
axis ratio of the surface brightness distribution measured with {\it
HST}.  We also used an {\it a priori} constraint on the amplitude of
the external shear ($\gamma=0.05\pm0.05$) based on limits from the
alignment between the major axes of lens models and observed lens galaxies 
(Kochanek~2002).  Ultimately we found that
these {\it a priori} constraints did not play a significant role in
determining the results.  The {\it a priori} constraint on the Hubble
constant was $H_0=72 \pm 7 $~km~s$^{-1}$~Mpc$^{-1}$. We first described G
with an ellipsoidal pseudo-Jaffe model [$\rho \propto
r^{-2}(r^2+a^2)^{-1}$]. In this model, the break radius $a$ can be
varied continuously, allowing the mass distribution to be adjusted
from the limit of a point mass ($a\rightarrow 0$) to the limit of an
isothermal mass distribution ($a\rightarrow\infty$). We described G22
as either a point mass, a singular isothermal sphere (SIS), a singular
isothermal ellipsoid (SIE), or a pseudo-Jaffe model.

\begin{figure}
\plottwo{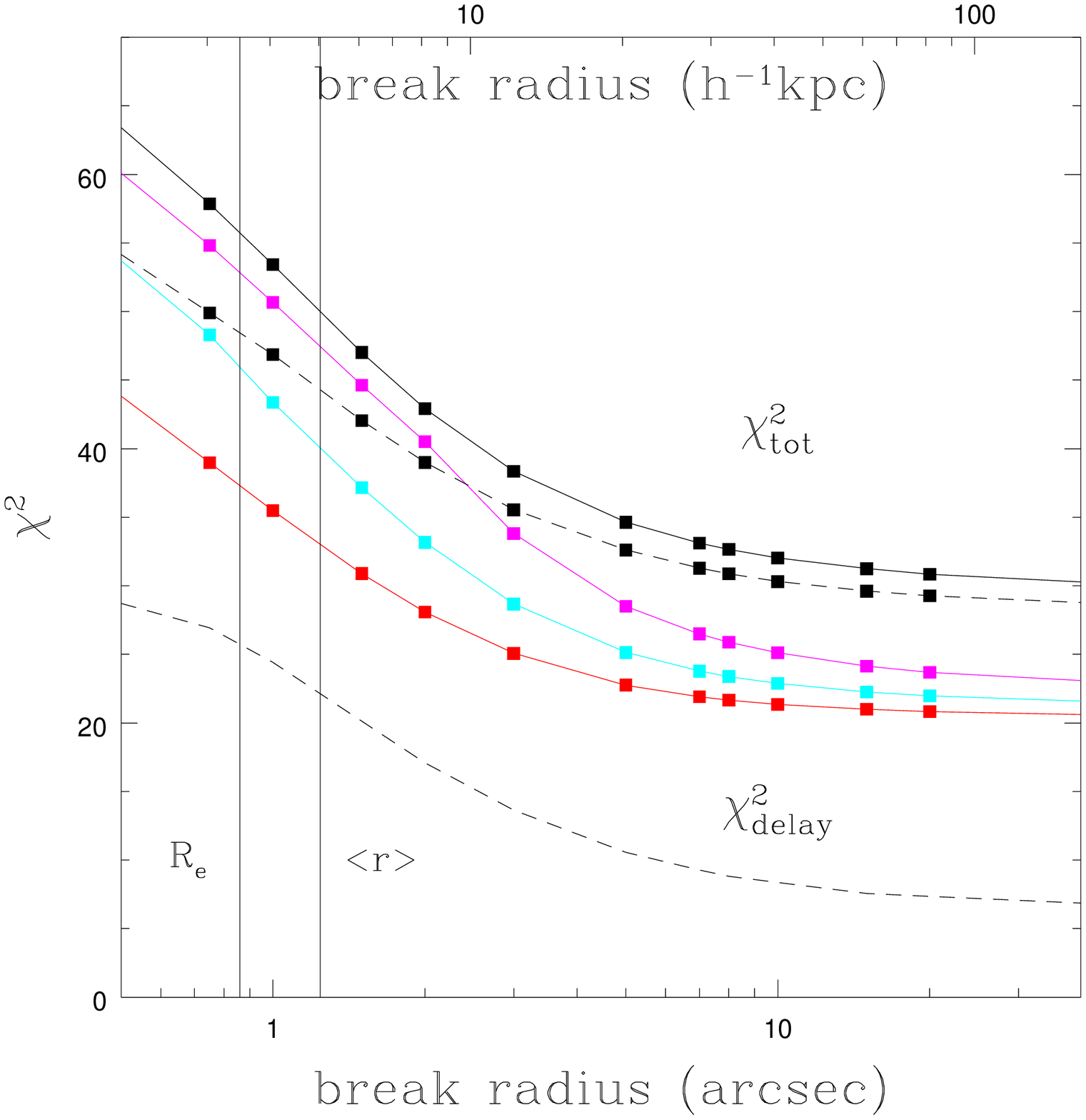}{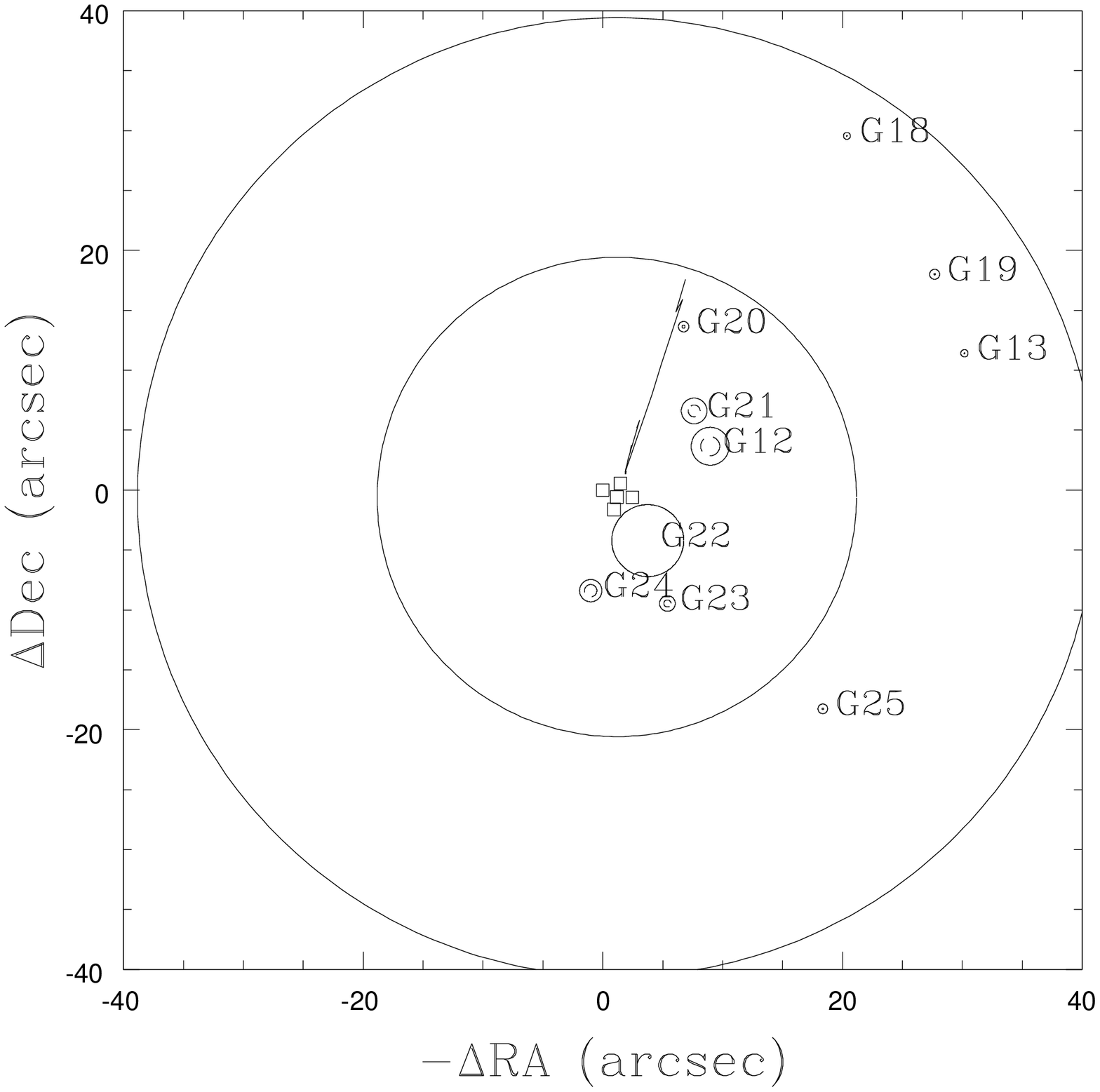}
\caption{(Left)
 The goodness-of-fit parameter as a function of $a$, the break radius
 of the pseudo-Jaffe model for the lens galaxy G.  The break radius is
 the length scale at which the rotation curve of the galaxy
 undergoes a transition from flat to Keplerian.  The uppermost solid
 curve is for a model with an external shear. The next three solid
 curves are for models in which the lens galaxy is embedded in a group
 NFW halo with a break radius of either $r_c=20\farcs0$, $10\farcs0$
 or $5\farcs0$ (from top to bottom).  The uppermost dashed curve shows
 the result when the group halo is modeled as an SIS.  The variation
 in $\chi^2$ is governed mainly by the constraint on the 
 observed time delays; this is illustrated by the lower dashed curve,
 which shows $\chi^2_{\rm delay}$ for the external shear model. The
 vertical lines mark the mean radius of the quasar images,
 and the effective radius of the lens galaxy.
 }
\label{fig:pjaffe}
\caption{(Right)
 The environment of \he0435. The 5 squares near the center of the map
 show the positions of the lens galaxy and the 4 quasar images. The
 large circles centered on the lens are the $20\farcs0$ and
 $40\farcs0$ radii, for visual reference. The small circles mark the
 locations of nearby galaxies that were identified by Morgan et
 al.~\cite{Morgan2004p00}, and the sizes of each circle are
 proportional to the expected amplitude of the lensing perturbation
 from that galaxy. Solid circles are for external shear, and dashed
 circles are for higher-order perturbations. The line between the lens
 galaxy and G20 is the locus of allowed positions for the centroid of
 a group halo.
 }
\label{fig:enviro}
\end{figure}

In the best-fitting model, the mass of the lens galaxy within the
cylinder bounded by the Einstein ring ($b=1\farcs18=4.8h^{-1}$~kpc) is
$2.2 \times 10^{11}h M_\odot$.  As always, this is the most robust
result of lens modeling, with a negligible statistical
uncertainty. The corresponding velocity dispersion for an SIS model is
$\sigma = 255$~km~s$^{-1}$. The best-fitting value of the break radius
is $a\rightarrow \infty$, so the best-fitting pseudo-Jaffe mass distribution for
G is essentially isothermal. The lower limit on the break radius is $a
> 14\farcs7=59h^{-1}$~kpc at 1$\sigma$ confidence, and $a >
5\farcs4=22h^{-1}$~kpc at 2$\sigma$ confidence. These constraints are
illustrated in Fig.~\ref{fig:pjaffe}, which shows how the
goodness-of-fit parameter ($\chi^2$) varies with the choice of break
radius.  (Also plotted in Fig.~\ref{fig:pjaffe} are the results from
the group-halo models discussed in \S~\ref{subsec:group-halo}.) The
requirement that the model match the observed time delays 
leads to the constraint on the break radius.

Notably, even in the best-fitting model, in which the mass
distribution has a nearly flat rotation curve, the fit is rather poor
($\chi^2_{\rm delay} \simeq 7$ for $a=40\farcs0$). One way to state
the difficulty is that the favored value of the Hubble constant is
$H_0\simeq 94$~km~s$^{-1}$~Mpc$^{-1}$, as compared to the constraint
of $H_0=72 \pm 7 $~km~s$^{-1}$~Mpc$^{-1}$. Since the time delays of
the model scale as $\Delta t \propto (1-\langle\kappa\rangle)/H_0$, it
follows that a better fit would be obtained if $\langle\kappa\rangle$
were allowed to be larger than 0.5, i.e., if the rotation curve of the
galaxy were rising rather than flat.

\subsection{The Effect of the Lens Galaxy Environment}
\label{subsec:environment}

Before considering further changes in the radial structure of the lens
galaxy, we explore the possible influences of the lens galaxy
environment on the measured quantities. Fig.~\ref{fig:enviro} is a map
of the lens galaxy and its closest neighbors, in which the symbol
sizes encode the relative strengths of the perturbations expected from
each galaxy.  In creating this map, we assumed that the Einstein
radius of each galaxy was proportional to the square root of its
I-band luminosity ($b_i \propto \sigma_i^2 \propto \sqrt{L_i}$). The
amplitude of the external shear is then expected to be proportional to
$(b_i/R_i)$, and higher-order perturbations are proportional to $(b_i
b /R_i^2)$, where $b$ is the Einstein radius of the lens galaxy and
$R_i$ is the angular separation from the lens galaxy. The dominant
perturbation comes from G22, as it is both nearby and fairly luminous.

\begin{figure}
\plottwo{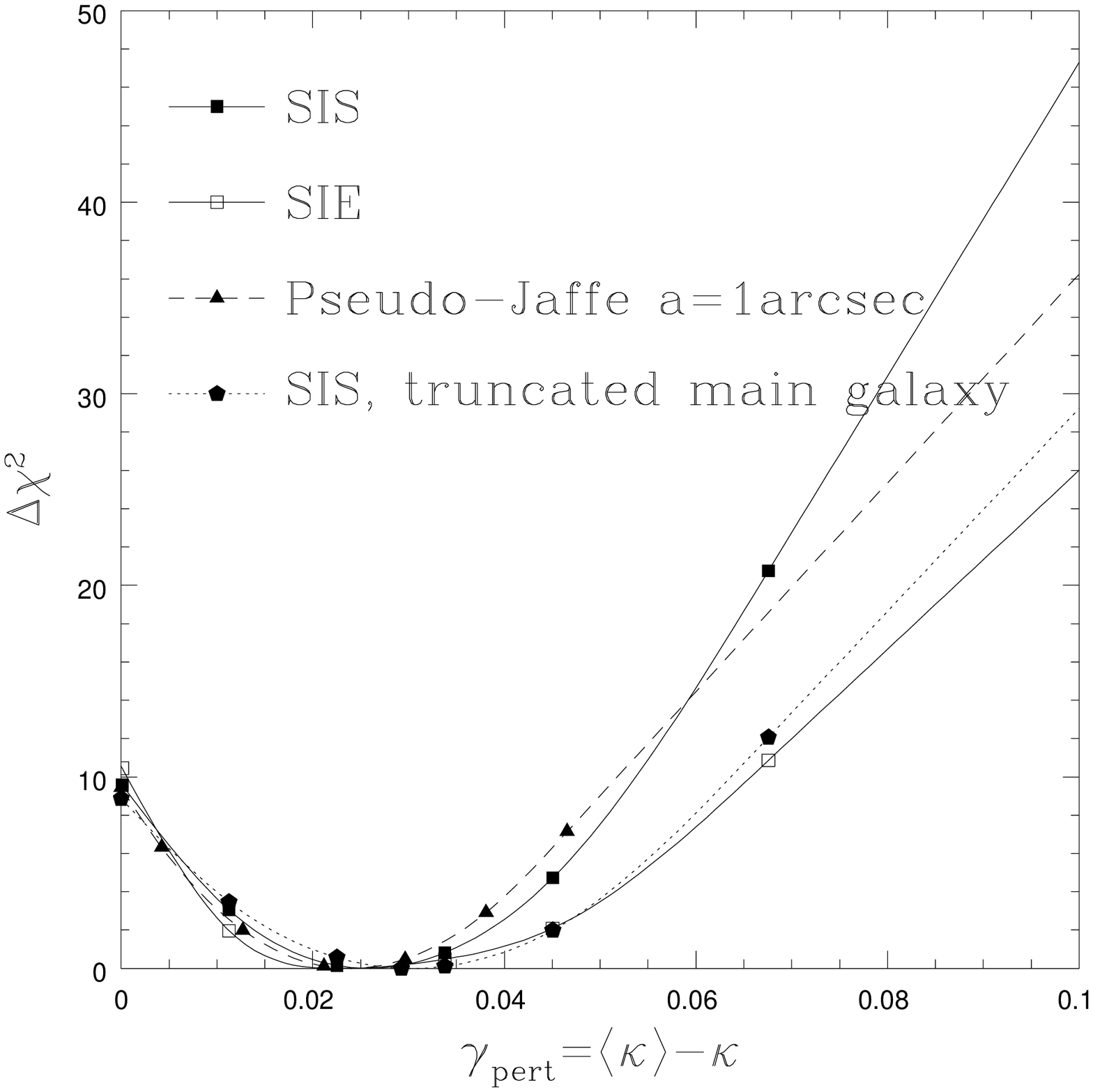}{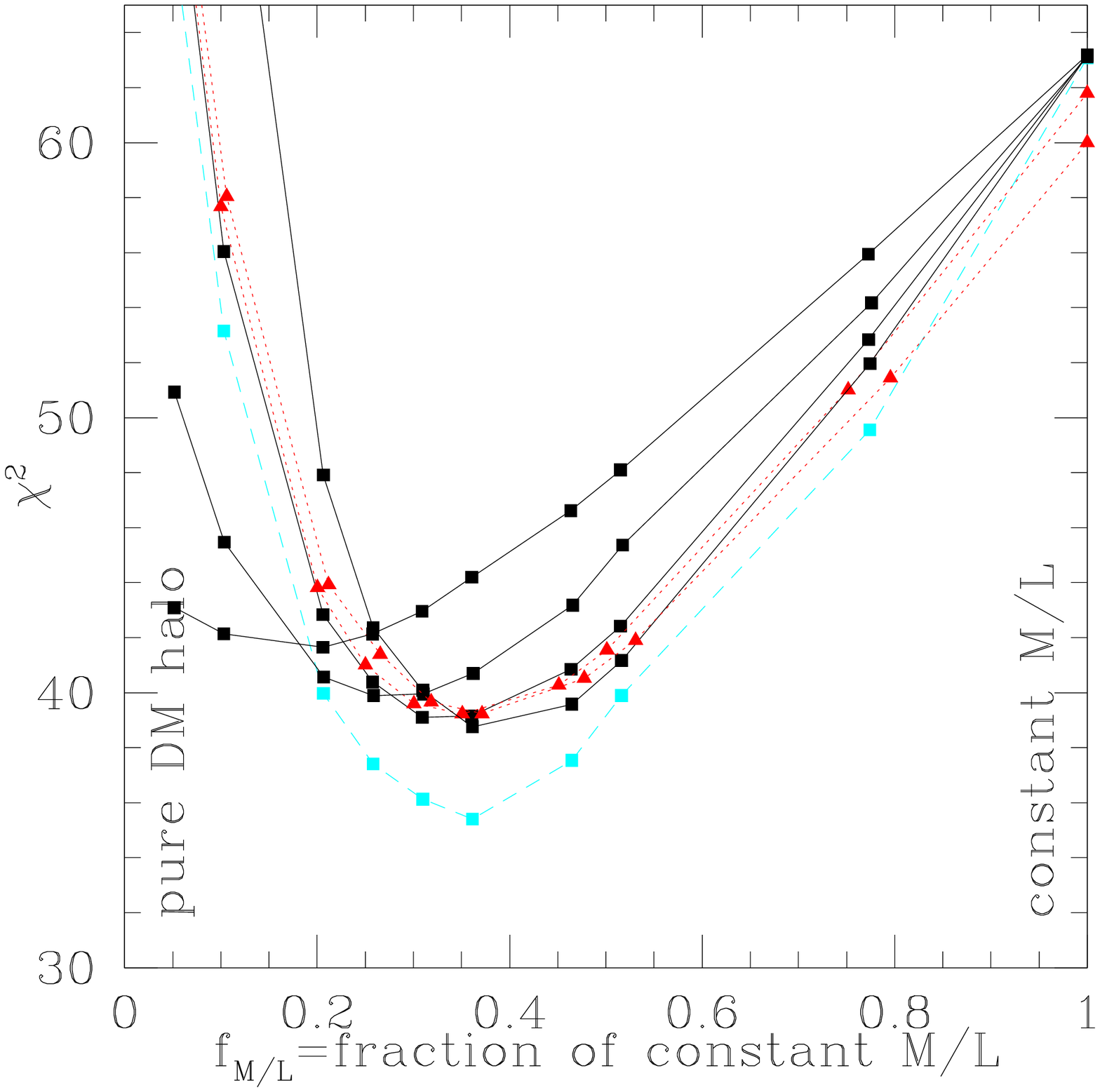}
\caption{ (Left)
 The goodness-of-fit parameter is plotted as a function of the shear
 $\gamma_{\rm pert}$ induced by the neighboring galaxy G22.  Although
 parametrized by the shear, the constraints are due to the higher 
 order perturbations produced by G22 -- the shear itself is degenerate
 with the global, external shear included as part of the model.  
 Results are
 shown for cases in which G22 is described by an SIS (solid squares),
 SIE (open squares), and a pseudo-Jaffe model with a break radius of
 $a=1\farcs0$ (dashed, filled triangles).  In all of these cases, the
 main lens galaxy is described as a pseudo-Jaffe model with
 $a=10\farcs0$.  If instead $a=2\farcs0$ for the main galaxy, the
 Einstein radius of the SIS model for G22 increases slightly (dotted,
 pentagons).
 }
\label{fig:neighbor}
\caption{ (Right)
 Results of ``galaxy plus halo'' (de~Vaucouleurs plus NFW) models. The
 goodness-of-fit parameter is plotted as a function of $f_{M/L}$,
 which is the mass of the de~Vaucouleurs component divided by the mass
 of the best-fitting pure-de~Vaucouleurs model. The 4 solid curves
 with squares show models with $R_e=0\farcs86$ and $r_c=2\farcs5$,
 $5\farcs0$, $10\farcs0$ and $20\farcs0$ (in increasing order of the
 optimum value of $f_{M/L}$). The dashed line with squares shows the
 results for the case in which $R_e=0\farcs86$ and the NFW halo shape
 is allowed to vary freely (as opposed to being constrained by the
 observed surface brightness distribution).  The dotted curves with
 triangles show the results for $R_e=1\farcs0$ and $1\farcs2$ with
 $r_c=10\farcs0$.
 }

\label{fig:nfw}
\end{figure}

We cannot detect the external shears produced by individual galaxies because
they are degenerate with the global external shear -- the effect of an
individual galaxy can be detected only through higher-order perturbations.  
As shown in Fig.~\ref{fig:neighbor}, the higher-order perturbations 
generated by G22 are measurable and correspond to a constraint on its
shear of $\gamma_{{\rm G}22}=0.025\pm0.008$,
regardless of whether G22 is modeled as an SIS or a point mass, and
regardless of the structure of the main lens galaxy G. When G22 is
modeled as an SIS, the best-fitting Einstein radius is
$0\farcs22\pm0\farcs07$, corresponding to a circular velocity $v_c =
157\pm 25$~km~s$^{-1}$. Since G22 should have negligible surface
density at projected radii corresponding to the 
$R_{{\rm G}22}\simeq 4\farcs4 \simeq 18h^{-1}$~kpc projected
separation of G22 and the lens, the shear that is produced by G22 equals the
mean surface density of interior to $R_{{\rm G}22}$.  In short, the mass of G22 can be
estimated from the higher order perturbations it produces, and the result is 
$M_{{\rm G}22}(R<R_{{\rm G}22})=(7.9\pm 2.6) \times 10^{10} hM_\odot$.

As discussed by Morgan al.~\cite{Morgan2004p00}, we can use this
measurement of the mass of G22 to determine whether or not this
companion galaxy possesses a dark matter halo.  Based on our 
estimate of the evolution-corrected B-band luminosity of G22
and the B-band Tully-Fisher relations of Kannappan, Fabricant \&
Franx~(\cite{Kannappan2002p2358}), G22 should have a circular
velocity of $v_c \simeq 190\pm 45$~km/s.  This is very close to 
the circular velocity of $v_c=150\pm 25$~km~s$^{-1}$ 
predicted for G22 assuming an SIS mass distribution and the
measured critical radius.  Since the lens model constrains
the mass $M_{{\rm G}22}(R<R_{{\rm G}22})$, the predicted
circular velocity depends only the scale length $R_d$ of G22's
mass distribution, $v_c^2 \propto M_{{\rm G}22}(R<R_{{\rm G}22})/R_d$.
As we make the mass distribution of G22 more compact (smaller $R_d$),
then we predict a higher circular velocity.  If we use our
best fit disk plus bulge model from the HST data as a constant-$M/L$
model for the mass distribution, then we predict $v_c \simeq 275$~km~s$^{-1}$,
while if we add a standard NFW halo normalized so that the stars
represent only 15\% of the mass, we predict $v_c\simeq 180$~km~s$^{-1}$.  
Thus, if the Tully-Fisher estimate of the
circular velocity is correct, G22 must have a significant 
dark matter halo.

Unfortunately, we find that this same technique cannot be used to
learn much about the other neighboring galaxies, because the
perturbations that they produce are too small.  The next-largest
perturbation comes from the close pair of galaxies G12 and G21, which
together should produce higher-order perturbations that are only 40\%
of strength of the effect of G22.  We computed a series of models in
which G21 was represented as an SIS, in addition to G22.  The result
was a small improvement in the fit, and a best-fitting Einstein radius
for G21 of $b_{{\rm G}12}=0\farcs17$ (or, equivalently, $\gamma_{{\rm
G}12}=0.008$ and $M_{{\rm G}12}(R<R_{{\rm G}12}) = 1.0 \times 10^{11}
h M_\odot$, where $R_{{\rm G}12}=8\farcs9=36h^{-1}$~kpc is the projected
separation of G12 and G). But the uncertainties are large enough
that at at $1\sigma$ confidence we can place only upper bounds of
$b_{{\rm G}12}< 0\farcs28$, $\gamma_{{\rm G}12} < 0.02$, and $M_{{\rm
G}12}(R<R_{{\rm G}12}) <2.5 \times 10^{11} h M_\odot$.  Given the
large uncertainties, it is fruitless to try to distinguish the effects
of G12 and G21 separately.

We also tried models that incorporated an {\it a priori} constraint on
the external shear, $\gamma = 0.00 \pm 0.01$, thereby requiring that
the angular structure of the lens potential be determined entirely by
the primary lens galaxy and the SIS model components representing neighboring galaxies. We
included SIS components for all the observed galaxies within
20\arcsec. We further assigned weak priors (50\% accuracy) that
encouraged the Einstein radii of the SIS models to agree with the
estimates based on the observed I-band luminosities. The resulting
fits to the data were only slightly poorer than the cases in which the
external shear was allowed to vary independently.  We conclude that
the nearby galaxies are sufficient to explain most of the angular
structure in the lens potential, but that there is a small component
($\gamma_{\rm ext}\simeq 0.01$ with a position angle of about
$-20^\circ$) that is not easily attributable to the nearby galaxies.
This is a reasonable result, in light of previous work that has shown
that large-scale structure along the line of sight should generally
provide a contribution to the shear that is of this order of magnitude
(e.g. Barkana~\cite{Barkana96}). Similar results were obtained when we
enlarged the sample of SIS galaxies to include all the observed
galaxies within 40\arcsec.

\subsection{The Possibility of a Group Halo}
\label{subsec:group-halo}

It is possible that the mass distribution is better described by a
single ``group halo'' rather than by individual SIS components at the
locations of all the observed galaxies.  However, models in which 
the centroid of the group halo do not coincide with the position
of a galaxy are not theoretically popular -- for example, models of
the halo occupancy distribution (HOD) argue that all halos have a
central galaxy and that the central galaxy is generally the most 
massive (e.g. Cooray \& Sheth~\cite{Cooray02}).  In
the models described in \S~\ref{subsec:parametric} (consisting of a
primary lens, the perturber G22, and an external shear) the external
shear has an amplitude of $\gamma \sim 0.05$ and a position angle of
approximately $-30^\circ$.  Morgan et al.~(2004) provided some
evidence that the centroid of the galaxy group is in that general
direction.  We explored for this possibility by replacing the
independent external shear parameter with a mass component
representing a group halo.  We tried both an SIS and an NFW mass
distribution to describe the group halo.  The results are plotted in
Fig.~\ref{fig:pjaffe} along with the previously described results for
the external-shear models.

Generally speaking, the group-halo models provide a better fit to the
data than the external-shear models.  Within the group-halo models,
those that have a higher convergence (for a fixed shear) produce
better fits. The extra convergence allows a better fit to the time
delays.  Consequently, an NFW halo is favored over an SIS halo, and
large NFW break radii are favored over small NFW break radii.
Regardless of the form of the group-halo model, the data still impose
a strong constraint on the break radius of the primary lens galaxy.
The $2\sigma$ lower limit on $a$ is $4\farcs9$ for an SIS group halo.
For an NFW group halo, the corresponding lower limits are $3\farcs4$,
$4\farcs7$, and $6\farcs4$, for an NFW break radius of $20\farcs0$,
$10\farcs0$, and $5\farcs0$, respectively.

The best-fitting centroid positions for the group halo do not seem to
be associated with any of the observed galaxies (see Fig.~\ref{fig:enviro}),
which may be unphysical.  The key property of the group-halo models that accounts for
their superior fit to the data is that they are allowed the freedom to
contribute to the local surface density near the primary lens galaxy,
unlike the external-shear models.  In fact, if there were a massive
group halo, one would expect the primary lens galaxy to lie at its
center, since the primary lens galaxy is the most luminous galaxy in
the field.  Its Einstein radius is nearly 5 times larger than that of
the second-place galaxy G22, and its luminosity is nearly 4 times that
of G22.

\subsection{Embedding the Lens Galaxy in a Dark Matter Halo}
\label{subsec:rotcurve}

All of the the preceding calculations have suggested that a
concordance between $H_0 = (72\pm 7)$~km~s$^{-1}$~Mpc~$^{-1}$ and the
measured time delays of \he0435\ requires that the primary lens galaxy
should have a rising rotation curve.  Even our SIS models for G did
not have a large enough surface density at the location of the
Einstein radius to be compatible with the accepted value of $H_0$.

In order to build a model with a rising rotation curve, we combined a
constant-$M/L$ model with an NFW halo.  For the constant-$M/L$
component (the ``galaxy'' or ``stellar component''), we considered
models in which the galaxy has an effective radius of $R_e=0\farcs86$,
$1\farcs0$ or $1\farcs2$.  We constrained its axis ratio to be
$0.79\pm0.04$, and the position angle of its major axis to be
$-5^\circ\pm4^\circ$, to match the observed surface brightness
distribution.  For the NFW component (``halo''), we considered four
different values of the break radius: $r_c=2\farcs5$, $5\farcs0$,
$10\farcs0$ and $20\farcs0$.  The halos were centered on the stellar
mass distribution and were generally constrained to have the same
ellipsoidal shape.  (Experiments in which the halo shape was allowed
complete freedom did result in slightly improved fits, but did not
change any of the conclusions described below about the radial mass
distribution.)  Since only G22 produces significant higher-order
perturbations beyond an external shear, we included an SIS component
representing G22.  The cumulative effect of all the other galaxies
(and large-scale structure) was represented by an external shear.

The results of these models are shown in Fig.~\ref{fig:nfw}.  The
variation in the goodness-of-fit parameter $\chi^2$ is plotted as a
function of $f_{M/L}=M_\star/M_{\star, {\rm no halo}}$. This
dimensionless factor is proportional to the mass of the stellar
component. It has been divided by the mass of the stellar component in
the best-fitting model with no NFW halo (a purely constant-$M/L$
model).  The $\chi^2$ values in this plot cannot be directly compared
to those in Fig.~\ref{fig:pjaffe}, because of the additional
constraints on the shape of the stellar mass distribution that we are
imposing in this case.

\begin{figure}
\plotone{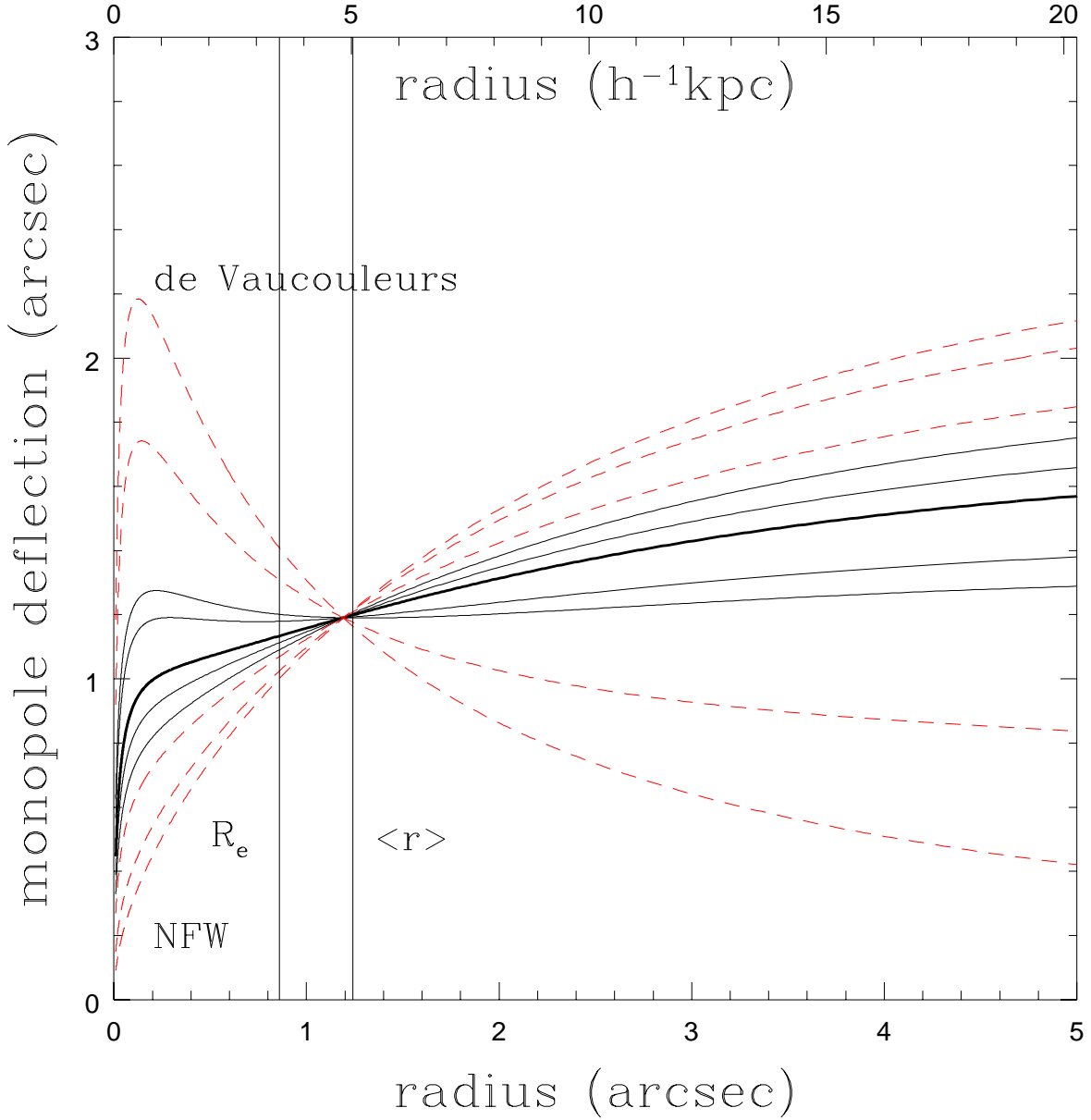}
\caption{
 Radial deflection profiles for the ``galaxy plus halo''
 (de~Vaucouleurs plus NFW) models, with $R_e=0\farcs86$ and
 $r_c=10\farcs0$.  The heavy solid curve is the best-fitting model and
 the light solid curves are models that also provide an acceptable fit
 ($\Delta\chi^2<4$).  The deflection profiles plotted here are
 approximately the square of the rotation curve.  The best fitting
 pseudo-Jaffe models have a constant deflection profile (flat rotation
 curve).  A constant-$M/L$ (de~Vaucouleurs only) model has a falling
 rotation curve and results in a poor fit to the data.  The vertical
 lines mark the effective radius $R_e$ and the mean radius of the
 images $\langle r\rangle$.
 }
\label{fig:deflect}
\end{figure}

The results can be summarized as follows:
\begin{enumerate}

\item Neither a constant-$M/L$ model, nor a pure-halo model, can
provide an acceptable fit to the data. These two extremes can be ruled
out with high confidence.

\item The best-fitting value of $f_{M/L}$ has a weak dependence upon
the scale lengths of the mass components.  For an effective radius of
$R_e=0\farcs86$, and NFW break radii of $r_c=2\farcs5$, $5\farcs0$,
$10\farcs0$, and $20\farcs0$, we find $f_{M/L}=0.18\pm0.10$,
$0.28\pm0.08$, $0.34\pm0.07$ and $0.39\pm0.07$, respectively.  This
dependence can be understood as the necessity for $f_{M/L}$ to
increase as the break radius increases, in order to maintain a fixed
surface density near the Einstein ring.  Similarly, if we hold
$r_c=10\farcs0$ fixed, and make the galaxy less centrally concentrated
with the choice $R_e=1\farcs0$, then $f_{M/L}$ increases
correspondingly to $0.36\pm0.08$.

\item All of the best-fitting models agree on the value of the surface
density within the annulus bounded by the images:
$\langle\kappa\rangle\simeq0.60\pm0.05$. As explained in
\S~\ref{subsec:analytic}, it is this quantity that controls the
connection between the observed time delays and the Hubble constant.
Note that this value of $\langle\kappa\rangle$ agrees well with the
simple analytic estimate of \S4.1, is 20\% larger than SIS
value of 0.5, and is much larger than the constant-$M/L$ value of
0.22.

\item If the effective radius of the stellar component is held fixed,
then $\langle\kappa_\star\rangle$ (the {\it stellar} surface density
at the Einstein radius) is correlated with the NFW break radius, in
the sense that more compact halos must be more dark-matter dominated.
For example, if $R_e=0\farcs86$, then $ \langle\kappa_*\rangle\simeq
0.05\log(h r_c/\hbox{kpc})$.
Again, this can be understood as the necessity to
maintain a sufficiently high total surface density near the locations
of the quasar images.

\item Fig.~\ref{fig:deflect} shows the deflection profiles of the
best-fitting models with both NFW and constant-$M/L$ components.  The
deflection profile is roughly equivalent to the square of the rotation
curve. At the location of the Einstein radius, the deflection profiles
are increasing functions of radius, i.e., the galaxy has a
slightly rising rotation curve.

\end{enumerate}

These results from the lens models are supported by estimates of the
mass-to-light ratio of the galaxy.  The mass of the galaxy within its
Einstein radius is determined from the measured redshifts and
astrometry, without appreciable systematic error.  Given the estimated
B-band luminosity of the galaxy (\S~\ref{sec:hst}), we can estimate
the mass-to-light ratio of all the material within the Einstein
radius.  For the measured value $R_e=0\farcs86$, the luminosity within
the Einstein radius is 58\% of the total luminosity. Putting it all
together, the implied B-band mass-to-light ratio within the Einstein
ring is $\Upsilon_B=15.6\pm1.9$ as observed, and
$\Upsilon_B=19.0\pm4.2$ when corrected for evolution to redshift zero.
(The uncertainties are dominated by the uncertainty in the
luminosity.)  Given the local estimate of $\Upsilon_*=7.8\pm2.7$ by
Gerhard et al.~\cite{Gerhard2001p1936}, the implication is that stars
represent only $(40\pm17)$\% of the mass projected inside the Einstein ring,
which is consistent with the results of the lens models.

\subsection{The Connection to Theoretical Halo Models}
\label{subsec:adiabatic}

In order to connect these models to theoretical expectations we must
relate the halo parameters to estimates from simulation and account
for the changes in the dark matter distribution produced by they
baryonic component known as ``adiabatic compression'' (Blumenthal et al.~\cite{Blumenthal86}).
In principle, we could use adiabatically compressed halos in the lens modeling, 
but in practice it would slow down the computations by an unacceptable
degree because of the need for ellipsoidal rather than circular lens models.  
Recall, however, that the data really only specify the mass within the Einstein ring 
and the surface density at the ring. With this in mind, we searched for 
adiabatically compressed halo models that match these key observables.
For each of the ``galaxy plus NFW'' models described in the previous
section, we calculated the mass within the Einstein radius and the
surface density at the Einstein radius.  As a matter of convenience,
we used a Hernquist model (Hernquist~\cite{Hernquist1990p359}) instead of the
de~Vaucouleurs model. Hernquist models have the desirable property
that the enclosed mass as a function of radius can be computed
analytically in both two and three dimensions.  We then searched a
collection of adiabatically-compressed halo models for examples that
satisfy the constraints on the enclosed mass and the local surface
density.  The collection of halo models was the same collection that
was computed previously by Kochanek~\cite{Kochanek2003p49}, which in
turn was based upon the early-type lens models of Keeton~\cite{Keeton2001p46}
and the Bullock et al.~\cite{bullock2001p559}
parameterizations of halo virial masses and concentrations.

\begin{figure}
\plottwo{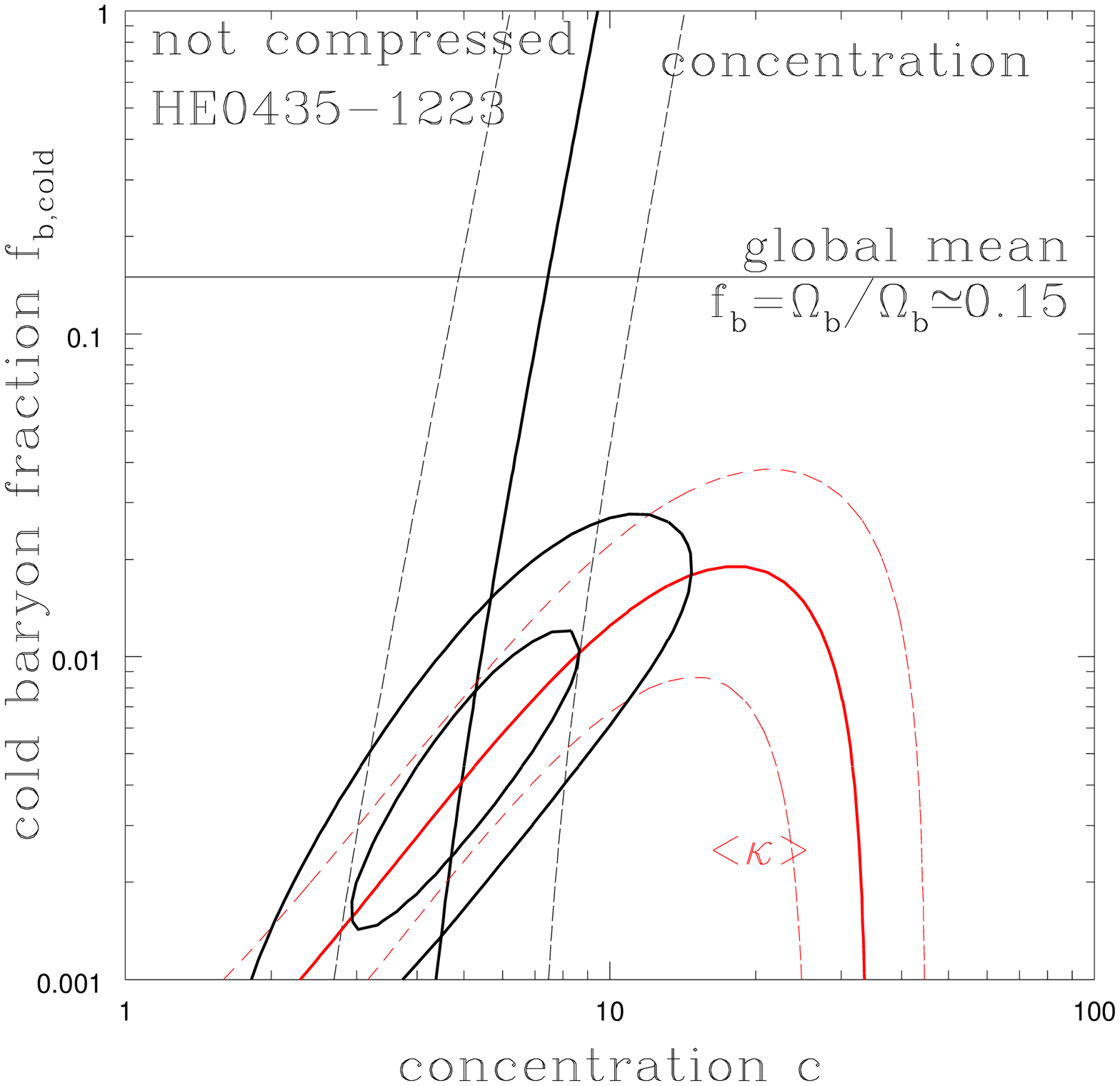}{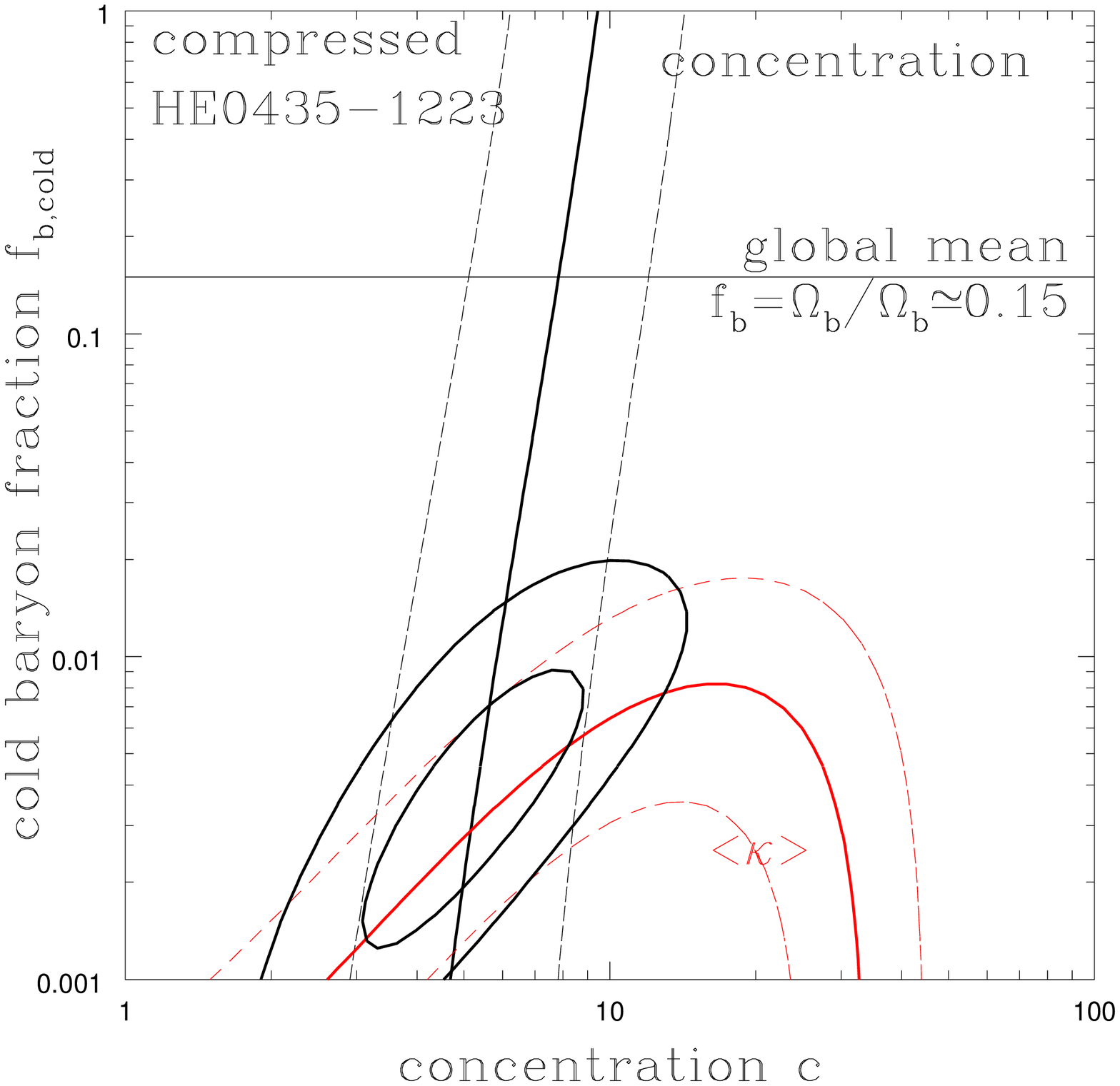}
\plottwo{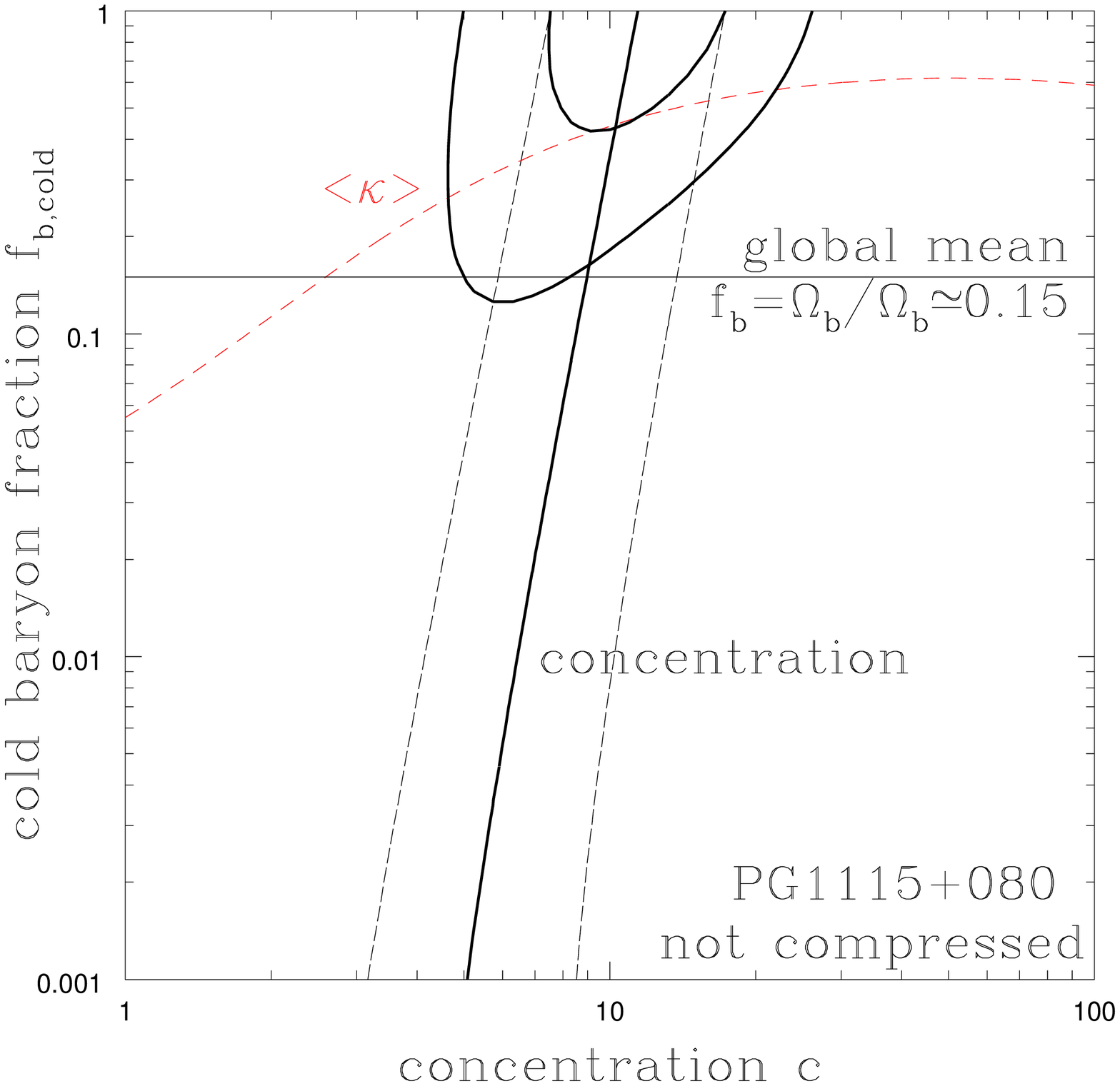}{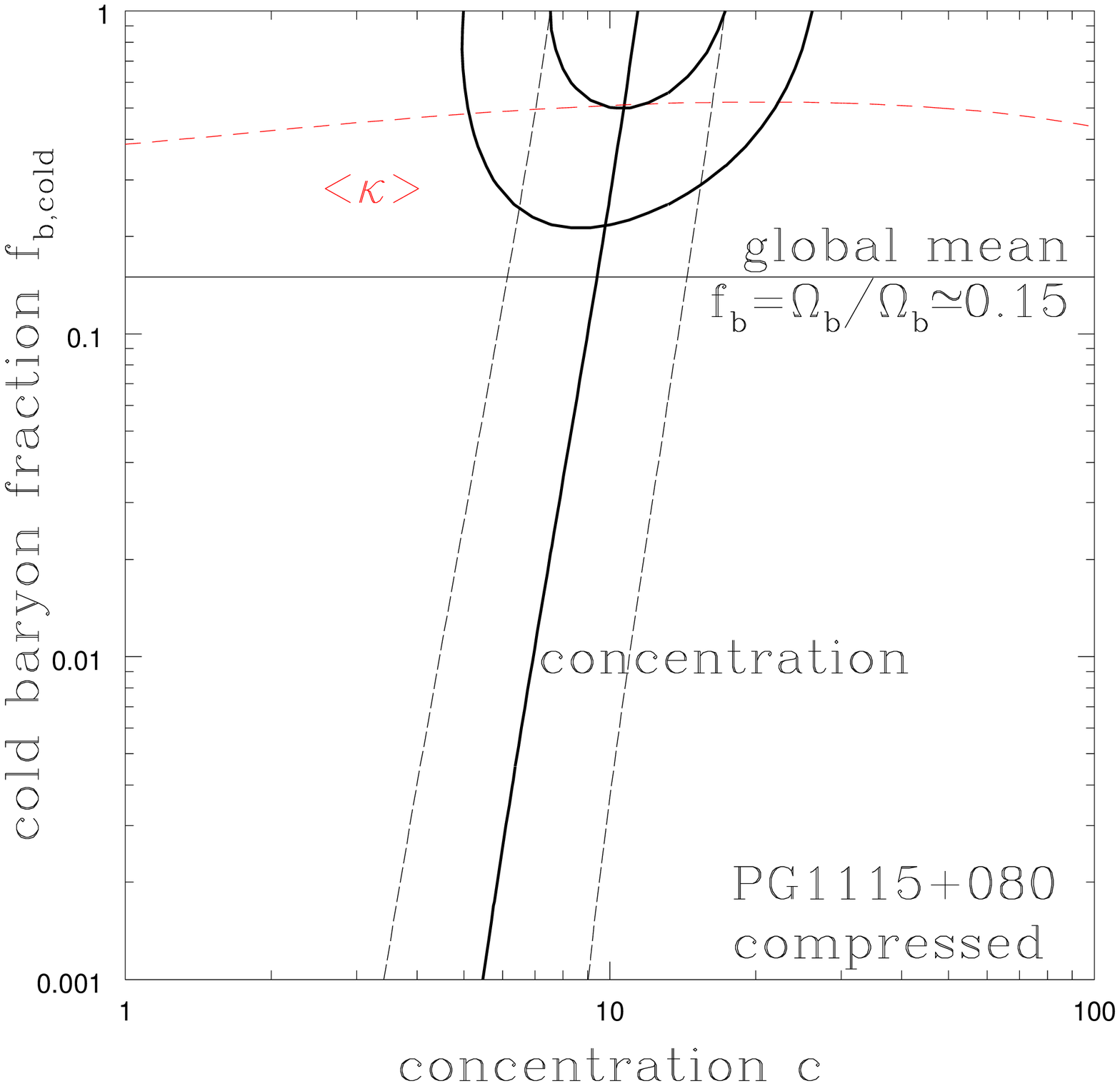}
\caption{
  Standard halo models
  for \he0435\ (top) and PG~1115+080 (bottom), without adiabatic
  compression (left) and with adiabatic compression (right).  The
  error ellipses show the 68\% and 95\% confidence regions in the
  space of the halo concentration $c$ and the cold baryon fraction
  $f_{b,{\rm cold}}$.  The constraint is essentially the product 
  of the constraint on the halo concentration for a given halo mass 
  (roughly vertical solid and dashed lines showing the mean and 
  typical dispersion in the concentration) and the surface
  density at the Einstein ring $\langle\kappa\rangle$ (the 
  second set of solid and dashed lines).  For PG1115+080, 
  only the lower bound on $\langle\kappa\rangle$ is visible.
  A horizontal line marks the global mean value for the 
  ratio between the baryonic and total masses of halos.
 }
\label{fig:model1}
\end{figure}

First, we consider halos without applying any adiabatic compression.
Fig.~\ref{fig:model1} shows the ability of these models to fit the
data, as a function of the halo concentration and the cold baryonic
mass fraction $f_{b,{\rm cold}}$. The latter quantity represents the
stellar component. In a constant-$M/L$ model, $f_{b,{\rm cold}}=1$,
but in a standard CDM halo it is limited to the global baryonic
mass fraction of $f_{b,{\rm cold}}<0.15=\Omega_b/\Omega_{\rm M}$ 
(Spergel et al.~\cite{Spergel2003p175}). The 
range of NFW break used in the models of \S4.5 span the range that i
s expected for a halos with the mass of the primary lens galaxy.  The models which
successfully fit the data have relatively low baryon fractions, $\log
f_{b,{\rm cold}}=-2.3\pm0.4$ compared to the global average, so only
$\sim 3\%$ of the baryons originally in the halo can have cooled to
make the observed galaxy.  The
best-fitting halos have a virial mass of $\log(M_{vir}/hM_\odot)=13.5
\pm 0.4$, a virial radius of $\log (h r_{vir}/\hbox{kpc})=2.7\pm0.1$,
and a break radius of $\log (h r_c/\hbox{kpc})=2.0\pm 0.3$.  The mass
of the stellar component is $\log (M_*/hM_\odot)=11.2\pm0.2$. The
contribution from dark matter is considerable even within the Einsten
radius.  In projection, the stars constitute $(40\pm 10)\%$ of the mass
inside the Einstein radius and $(30\pm 8)\%$ of the mass within two
effective radii.  In three dimensions, the stars constitute
$(74\pm11)\%$ of the mass inside a sphere with a radius of
$R_e$. Thus, measurements of stellar dynamics through optical
spectroscopy would be mainly sensitive to the stellar mass rather than
the dark matter.

Next, we include the effects of adiabatic compression.  The dark
matter becomes more important, and the inferred stellar surface
density must be adjusted to compensate.  We express this by
parameterizing the results as a function of the ratio of stellar
surface densities found with and without adiabatic compression.
Fig.~\ref{fig:model1} also shows the effect of adding adiabatic
compression to the models. The cold baryon fraction $\log f_{b,{\rm
cold}}=-2.4\pm0.4$ and the stellar mass $\log
(M_*/hM_\odot)=10.9\pm0.2$ are a factor of two smaller, but the virial
mass, virial radius, and break radius are little changed.  As a
result, the dark-matter fraction in the inner regions rises
appreciably, to $(77\pm9)\%$ and $(82\pm10)\%$ of the projected mass
inside the Einstein radius and $2R_e$, and to $(58\pm8)\%$ of the mass
inside a sphere of radius $R_e$.  

\section{Conclusions}
\label{sec:conclusions}

Our analysis of \he0435\ adds to the evidence that the structures of
early-type galaxy halos are heterogeneous, rather than homogeneous.
Assuming that $H_0=(72\pm 7)$~km~s$^{-1}$~Mpc$^{-1}$, most 
time-delay lens galaxies must have centrally
concentrated mass distributions that are strongly dominated by the
stars in their central regions (Kochanek 2002).  They must have {\it
falling} rotations curves and low surface densities at the
galactocentric distances where the lensed images occur (1--2$R_e$).
HE~0435--1223, by contrast, must have a rising rotation curve
at the radius of the lensed images.
As an illustration of this heterogeneity, we performed the same
``galaxy plus NFW'' analysis for the quadruple-image quasar
PG~1115+080. In that case, the best-fitting model has constant
mass-to-light ratio ($f_{M/L}>0.9$), a very small surface density at
the Einstein radius ($\langle\kappa\rangle=0.16\pm0.05$), and a
comparable stellar surface density.  If we focus on the
adiabatically-compressed model, then the estimated stellar mass in the
PG~1115+080 model [$\log (M_*/hM_\odot)=10.7\pm0.1$] is similar to the
estimate for \he0435, but the virial mass
[$\log(M_{vir}/hM_\odot)=10.9 \pm 0.1$] is far smaller. This is
because the cold-baryon fraction in the PG~1115+080 model is nearly
unity. In fact, the cold-baryon fraction must be {\it larger} that the
cosmological limit of $\Omega_b/\Omega_{\rm M} \simeq 0.15$.

The enormous difference between these halos is difficult to attribute
to any errors in our measurements of \he0435.  Our estimates for the
surface densities can be incorrect only in the event of a gross error
in the time delay measurements, or if there are contributions to
$\langle\kappa\rangle$ besides the primary lens galaxy and its
halo. The former possibility is very unlikely because of the multiple
correlated features that we observed in the light curves of the four
images.  The latter possibility is also unlikely, given the results of
the extensive suite of models presented in \S~\ref{sec:models-macro}.
For example, models with $f_{b,{\rm cool}}\simeq 0.1$ have annular
surface densities of $\langle\kappa\rangle \simeq 0.3$ rather than
$0.6$, which would require a time delay of $\Delta t_{AD}\simeq
-25$~days.  This is ruled out at very high confidence. We attempted to
find successful models in which some of the convergence was generated
by a nearby group halo, but we found that these models cannot remove
the requirement that the lens galaxy has a rising rotation curve; they
produce too much shear.

A natural context in which this heterogeneity of halos might be
understood is the halo model for populating dark matter halos with
galaxies (see the review by Cooray \& Sheth~\cite{Cooray02}). The typical early-type lens
galaxy should be a member of a group of galaxies. In the halo model,
one of the galaxies in a group lies at the center of the massive group
halo, while the other galaxies are smaller satellites orbiting in the
group halo. Because baryonic cooling enormously increases the central
densities of galaxies as compared to a pure dark-matter halo, the
lensing cross section is dominated by the individual cross sections of
the member galaxies, rather than the group as a whole (e.g. Kochanek
\& White~\cite{Kochanek2001p531}). However, it is
likely that the surface densities and stellar mass fractions of the
central galaxy and the satellite galaxies on scales larger than an
effective radius are very different.

In this context, one possible interpretation is that \he0435\ is the
central galaxy of its group.  It has a high dark-matter surface
density, and the stars constitute only a small fraction of the overall
baryonic mass of the group.  At the other extreme, the lens galaxy in
PG~1115+080 is a satellite galaxy with a partially stripped halo
all of whose baryons have been converted to stars.  The galaxy
G22 in \he0435\ is intermediate to these extremes.

Which type of galaxy dominates the lens population: central galaxies,
or satellite galaxies? The theoretical expectation is unclear.  It
depends on the detailed balance between the higher cross-section of
the more massive central galaxies and the large number of satellite
galaxies.  The present sample of time delay lenses suggests the two
populations are comparable.  The three lenses PG~1115+080,
HE~2149--2745 and B1600+434 require low dark-matter surface densities
(for $H_0=72$~km~s$^{-1}$~Mpc$^{-1}$) and are probably satellite
galaxies.  In contrast, \he0435\ and HE~1104--1805 require high
dark-matter surface densities, and are probably central galaxies of
groups.  This explanation for the heterogeneity of the time delay
lenses predicts that for time delay lenses that are satellite
galaxies, there should be a nearby, group center galaxy that has a
much higher mass-to-light ratio than either the lens or other
satellite galaxies.

\acknowledgements We thank E. Turner for organizing the APO observations,
J. Barentine and R. McMillan for conducting APO observations, and
P.\ Martini for occasionally observing on our behalf. This research made extensive use
of a Beowulf computer cluster obtained through the Cluster Ohio
program of the Ohio Supercomputer Center.  Work by J.N.W.\ was
supported by NASA through Hubble Fellowship grant HST-HF-01180.02-A,
awarded by the Space Telescope Science Institute, which is operated by
the Association of Universities for Research in Astronomy, Inc., for
NASA, under contract NAS~5-26555. Support for program HST-GO-9744 was
provided by NASA through a grant from the Space Telescope Science
Institute, which is operated by the Association of Universities for
Research in Astronomy, Inc., under NASA contract NAS~5-26666.

\begin{deluxetable}{ccc}
\tablecaption{Astrometry of \he0435}
\tablewidth{0pt}
\tablehead{ Component & $\Delta\hbox{RA} $ &$\Delta\hbox{Dec}$ }
\startdata
A  &$\equiv 0$ &$\equiv 0$  \\
B  &$+1\farcs476\pm0\farcs003$ &$+0\farcs553\pm0\farcs001$ \\
C  &$+2\farcs467\pm0\farcs002$ &$-0\farcs603\pm0\farcs004$ \\
D  &$+0\farcs939\pm0\farcs002$ &$-1\farcs614\pm0\farcs001$ \\
G  &$+1\farcs165\pm0\farcs002$ &$-0\farcs573\pm0\farcs002$ \\
\enddata
\label{tab:astrom}
\tablecomments{ For the NICMOS images we adopted pixel scales of
$\Delta x=0\farcs0760$ and $\Delta y = 0\farcs0753$.  }
\end{deluxetable}

\begin{deluxetable}{crrrcccc}
\scriptsize
\tablecaption{Photometry of \he0435}
\tablewidth{0pt}
\tablehead{ 
   Comp &H=F160W  &I=F814W  &V=F555W &$R_e$ &$\mu_e$        &$q$ &$\theta_{maj}$ \\
        &(mag)    &(mag)    &(mag)   &(")   &mag/arcsec$^2$ &    &(PA, deg) }
\startdata		    	      
A &$17.31\pm0.02$ &$17.84\pm0.02$ &$18.41\pm0.03$  \\
B &$17.80\pm0.02$ &$18.39\pm0.04$ &$18.99\pm0.07$  \\
C &$17.80\pm0.03$ &$18.41\pm0.02$ &$19.07\pm0.06$  \\
D &$18.06\pm0.02$ &$18.62\pm0.04$ &$19.12\pm0.04$  \\
G &$16.86\pm0.04$ &$18.85\pm0.02$ &$21.55\pm0.13$  &$0\farcs86\pm0\farcs04$ &$18.26\pm0.07$ &$0.79\pm0.04$ &$-5\pm4$ \\
H &$20.55\pm0.16$ &$22.06\pm0.17$ &$23.27\pm0.59$  &$0\farcs18\pm0\farcs02$ &               &$0.55\pm0.03$ &$-82\pm4$\\
\enddata
\label{tab:photom}
\tablecomments{ Magnitudes are given in the Vega system.  Components
A-D are quasar images, G is the primary lens galaxy, and H is the
quasar host galaxy. }
\end{deluxetable}

\begin{deluxetable}{crrcc}
\tablecaption{Comparison Stars in the Field of \he0435}
\tablewidth{0pt}
\tablehead{Star  &\multicolumn{1}{c}{$\Delta \hbox{RA}$}
                 &\multicolumn{1}{c}{$\Delta \hbox{Dec}$}
                 &\multicolumn{1}{c}{$F_0$}
                 &\multicolumn{1}{c}{$F_{obs}$}
                 }
\startdata
S1 &$-81\farcs$ &$ 48\farcs$ &$1.000\pm0.0100$ &$1.0046\pm0.0049$ \\
S2 &$ -9\farcs$ &$ 49\farcs$ &$1.117\pm0.0100$ &$1.1129\pm0.0043$ \\
S3 &$ -6\farcs$ &$-37\farcs$ &$0.966\pm0.0100$ &$0.9637\pm0.0037$ \\
S4 &$-66\farcs$ &$ 29\farcs$ &$0.893\pm0.0100$ &$0.8956\pm0.0047$ \\
S5 &$-28\farcs$ &$-38\farcs$ &$0.680\pm0.0100$ &$0.6792\pm0.0040$ \\
\enddata
\tablecomments{The relative positions given in this table are measured
in arc seconds east and north of quasar image A. $F_0$ is the defined
flux of the star used for the flux calibration and its prior
uncertainty. $F_{obs}$ is the mean flux of the star in the calibrated
SMARTS observations. The APO r-band fluxes were multiplied by $1.032$
to match the SMARTS R-band flux scale.}
\label{tab:stars}
\end{deluxetable}

\def\hm{\hphantom{-}}
\begin{deluxetable}{cccccccc}
\rotate
\scriptsize
\tablecaption{HE0435--1223 Lightcurves}
\tablewidth{0pt}
\tablehead{ HJD &\multicolumn{1}{c}{$\chi^2/N_{dof}$}
                &\multicolumn{1}{c}{QSO A} &\multicolumn{1}{c}{QSO B} 
                &\multicolumn{1}{c}{QSO C} &\multicolumn{1}{c}{QSO D}
                &\multicolumn{1}{c}{$\langle\hbox{Stars}\rangle$}
                &\multicolumn{1}{c}{Source} 
              }
\startdata
$2863.883$ &$  0.67$ &$ 2.065\pm 0.013$ &$ 2.632\pm 0.019$ &$ 2.592\pm 0.018$ &$ 2.821\pm 0.022$ &$-0.002\pm 0.003$ &SMARTS \\ 
$2871.829$ &$  1.67$ &$ 2.140\pm 0.015$ &$ 2.628\pm 0.022$ &$ 2.561\pm 0.020$ &$ 2.790\pm 0.023$ &$-0.003\pm 0.003$ &SMARTS \\ 
$2877.843$ &$  0.82$ &$ 2.134\pm 0.012$ &$ 2.644\pm 0.017$ &$ 2.670\pm 0.017$ &$ 2.803\pm 0.018$ &$\hm 0.002\pm 0.003$ &SMARTS \\ 
$2884.784$ &$  1.38$ &$ 2.141\pm 0.012$ &$ 2.653\pm 0.018$ &$ 2.661\pm 0.017$ &$ 2.870\pm 0.019$ &$\hm 0.001\pm 0.003$ &SMARTS \\ 
$2891.827$ &$  1.03$ &$ 2.172\pm 0.015$ &$ 2.689\pm 0.024$ &$ 2.685\pm 0.022$ &$ 2.831\pm 0.024$ &$-0.003\pm 0.003$ &SMARTS \\ 
$2899.843$ &$  0.93$ &$ 2.171\pm 0.012$ &$ 2.728\pm 0.018$ &$ 2.712\pm 0.017$ &$ 2.856\pm 0.019$ &$\hm 0.001\pm 0.003$ &SMARTS \\ 
$2906.848$ &$  2.25$ &$ 2.184\pm 0.009$ &$ 2.729\pm 0.011$ &$ 2.745\pm 0.011$ &$ 2.891\pm 0.012$ &$\hm 0.007\pm 0.002$ &SMARTS \\ 
$2916.803$ &$  1.56$ &$ 2.192\pm 0.010$ &$ 2.720\pm 0.015$ &$ 2.713\pm 0.014$ &$ 2.953\pm 0.016$ &$\hm 0.006\pm 0.002$ &SMARTS \\ 
$2919.828$ &$  0.65$ &$ 2.177\pm 0.013$ &$ 2.759\pm 0.020$ &$ 2.738\pm 0.019$ &$ 2.910\pm 0.021$ &$\hm 0.001\pm 0.003$ &SMARTS \\ 
$2926.780$ &$  3.08$ &$ 2.186\pm 0.012$ &$ 2.736\pm 0.018$ &$ 2.643\pm 0.016$ &$ 2.885\pm 0.019$ &$\hm 0.000\pm 0.003$ &SMARTS \\ 
$2937.672$ &$  2.54$ &$ 2.158\pm 0.010$ &$ 2.727\pm 0.015$ &$ 2.710\pm 0.014$ &$ 2.930\pm 0.016$ &$\hm 0.004\pm 0.002$ &SMARTS \\ 
$2941.689$ &$  2.91$ &$ 2.210\pm 0.010$ &$ 2.728\pm 0.013$ &$ 2.698\pm 0.012$ &$ 2.896\pm 0.014$ &$\hm 0.006\pm 0.002$ &SMARTS \\ 
$2947.698$ &$  1.03$ &$ 2.184\pm 0.011$ &$ 2.756\pm 0.016$ &$ 2.720\pm 0.016$ &$ 2.931\pm 0.018$ &$\hm 0.002\pm 0.002$ &SMARTS \\ 
$2954.711$ &$  0.75$ &$ 2.222\pm 0.021$ &$ 2.826\pm 0.037$ &$ 2.815\pm 0.035$ &$ 2.896\pm 0.037$ &$-0.012\pm 0.003$ &SMARTS \\ 
$2958.742$ &$  0.98$ &$ 2.179\pm 0.012$ &$ 2.791\pm 0.018$ &$ 2.772\pm 0.017$ &$ 2.875\pm 0.018$ &$\hm 0.003\pm 0.002$ &SMARTS \\ 
$2961.682$ &$  0.83$ &$ 2.204\pm 0.011$ &$ 2.746\pm 0.017$ &$ 2.772\pm 0.016$ &$ 2.952\pm 0.018$ &$\hm 0.005\pm 0.002$ &SMARTS \\ 
$2962.680$ &$  1.63$ &$ 2.196\pm 0.010$ &$ 2.734\pm 0.014$ &$ 2.740\pm 0.014$ &$ 2.934\pm 0.016$ &$\hm 0.004\pm 0.002$ &SMARTS \\ 
$2965.630$ &$  2.47$ &$ 2.160\pm 0.012$ &$ 2.744\pm 0.019$ &$ 2.809\pm 0.019$ &$ 2.965\pm 0.021$ &$\hm 0.000\pm 0.003$ &SMARTS \\ 
$2967.637$ &$  2.01$ &$ 2.177\pm 0.010$ &$ 2.773\pm 0.016$ &$ 2.723\pm 0.014$ &$ 2.936\pm 0.016$ &$\hm 0.006\pm 0.002$ &SMARTS \\ 
$2972.619$ &$  2.49$ &$ 2.215\pm 0.010$ &$ 2.770\pm 0.014$ &$ 2.763\pm 0.013$ &$ 2.997\pm 0.015$ &$\hm 0.009\pm 0.002$ &SMARTS \\ 
$2975.669$ &$  1.86$ &$ 2.174\pm 0.010$ &$ 2.775\pm 0.014$ &$ 2.728\pm 0.013$ &$ 2.919\pm 0.015$ &$\hm 0.005\pm 0.002$ &SMARTS \\ 
$2979.632$ &$  1.36$ &$ 2.176\pm 0.012$ &$ 2.690\pm 0.019$ &$ 2.758\pm 0.019$ &$ 2.947\pm 0.022$ &$\hm 0.000\pm 0.003$ &SMARTS \\ 
$2982.682$ &$  0.97$ &$ 2.154\pm 0.018$ &$ 2.839\pm 0.032$ &$( 2.847\pm 0.032)$ &$ 2.950\pm 0.035$ &$-0.010\pm 0.003$ &SMARTS \\ 
$2986.634$ &$  1.78$ &$ 2.166\pm 0.009$ &$ 2.722\pm 0.011$ &$ 2.723\pm 0.011$ &$ 2.949\pm 0.013$ &$\hm 0.006\pm 0.002$ &SMARTS \\ 
$2993.664$ &$  2.62$ &$ 2.143\pm 0.009$ &$ 2.724\pm 0.013$ &$ 2.721\pm 0.013$ &$ 2.916\pm 0.014$ &$\hm 0.005\pm 0.002$ &SMARTS \\ 
$3000.684$ &$  3.01$ &$ 2.146\pm 0.009$ &$ 2.701\pm 0.011$ &$ 2.715\pm 0.011$ &$ 2.923\pm 0.013$ &$\hm 0.006\pm 0.002$ &SMARTS \\ 
$3013.617$ &$  0.84$ &$ 2.136\pm 0.010$ &$ 2.694\pm 0.015$ &$ 2.714\pm 0.015$ &$ 2.902\pm 0.017$ &$\hm 0.003\pm 0.002$ &SMARTS \\ 
$3021.612$ &$  1.27$ &$ 2.108\pm 0.011$ &$ 2.734\pm 0.018$ &$ 2.744\pm 0.016$ &$ 2.895\pm 0.018$ &$\hm 0.005\pm 0.002$ &SMARTS \\ 
$3028.619$ &$  1.66$ &$ 2.097\pm 0.009$ &$ 2.704\pm 0.013$ &$ 2.704\pm 0.012$ &$ 2.859\pm 0.014$ &$\hm 0.006\pm 0.002$ &SMARTS \\ 
$3035.614$ &$  0.92$ &$ 2.082\pm 0.011$ &$ 2.715\pm 0.018$ &$ 2.694\pm 0.017$ &$ 2.846\pm 0.019$ &$\hm 0.001\pm 0.003$ &SMARTS \\ 
$3038.592$ &$  1.22$ &$ 2.077\pm 0.010$ &$ 2.670\pm 0.015$ &$ 2.676\pm 0.014$ &$ 2.842\pm 0.016$ &$\hm 0.003\pm 0.002$ &SMARTS \\ 
$3041.618$ &$  0.83$ &$ 2.093\pm 0.013$ &$ 2.706\pm 0.022$ &$ 2.679\pm 0.021$ &$ 2.859\pm 0.024$ &$-0.003\pm 0.003$ &SMARTS \\ 
$3045.653$ &$  0.89$ &$ 2.055\pm 0.017$ &$ 2.689\pm 0.030$ &$ 2.780\pm 0.029$ &$ 2.811\pm 0.030$ &$-0.007\pm 0.003$ &SMARTS \\ 
$3048.616$ &$  4.80$ &$ 2.096\pm 0.009$ &$ 2.719\pm 0.012$ &$ 2.735\pm 0.011$ &$ 2.809\pm 0.012$ &$\hm 0.007\pm 0.002$ &SMARTS \\ 
$3053.581$ &$  3.86$ &$ 2.120\pm 0.009$ &$ 2.715\pm 0.011$ &$ 2.737\pm 0.011$ &$ 2.819\pm 0.012$ &$\hm 0.007\pm 0.002$ &SMARTS \\ 
$3054.576$ &$  3.75$ &$ 2.100\pm 0.009$ &$ 2.749\pm 0.013$ &$ 2.758\pm 0.013$ &$ 2.793\pm 0.013$ &$\hm 0.006\pm 0.002$ &SMARTS \\ 
$3057.528$ &$  1.01$ &$ 2.137\pm 0.011$ &$ 2.722\pm 0.017$ &$ 2.749\pm 0.016$ &$ 2.831\pm 0.017$ &$\hm 0.003\pm 0.002$ &SMARTS \\ 
$3061.539$ &$  1.35$ &$ 2.138\pm 0.012$ &$ 2.737\pm 0.019$ &$ 2.749\pm 0.018$ &$ 2.850\pm 0.019$ &$\hm 0.001\pm 0.003$ &SMARTS \\ 
$3066.577$ &$  0.85$ &$ 2.102\pm 0.015$ &$ 2.808\pm 0.027$ &$ 2.803\pm 0.025$ &$ 2.852\pm 0.025$ &$-0.003\pm 0.003$ &SMARTS \\ 
$3067.597$ &$  1.33$ &$ 2.128\pm 0.012$ &$ 2.745\pm 0.019$ &$ 2.743\pm 0.018$ &$ 2.863\pm 0.020$ &$\hm 0.000\pm 0.003$ &SMARTS \\ 
$3069.533$ &$  1.78$ &$ 2.157\pm 0.013$ &$ 2.725\pm 0.020$ &$ 2.750\pm 0.019$ &$ 2.870\pm 0.021$ &$-0.003\pm 0.003$ &SMARTS \\ 
$3073.533$ &$  0.80$ &$ 2.160\pm 0.013$ &$ 2.761\pm 0.020$ &$ 2.727\pm 0.018$ &$ 2.900\pm 0.021$ &$\hm 0.000\pm 0.003$ &SMARTS \\ 
$3076.505$ &$  1.21$ &$ 2.166\pm 0.012$ &$ 2.762\pm 0.018$ &$ 2.757\pm 0.018$ &$ 2.874\pm 0.019$ &$\hm 0.000\pm 0.003$ &SMARTS \\ 
$3081.553$ &$  0.95$ &$ 2.129\pm 0.013$ &$ 2.748\pm 0.021$ &$ 2.779\pm 0.020$ &$ 2.912\pm 0.021$ &$\hm 0.000\pm 0.003$ &SMARTS \\ 
$3087.494$ &$  0.60$ &$ 2.173\pm 0.013$ &$ 2.795\pm 0.021$ &$ 2.766\pm 0.019$ &$ 2.897\pm 0.021$ &$\hm 0.001\pm 0.003$ &SMARTS \\ 
$3091.500$ &$  1.75$ &$ 2.162\pm 0.011$ &$ 2.795\pm 0.017$ &$ 2.797\pm 0.017$ &$ 2.884\pm 0.018$ &$\hm 0.002\pm 0.002$ &SMARTS \\ 
$3098.511$ &$  0.95$ &$ 2.179\pm 0.016$ &$ 2.829\pm 0.028$ &$ 2.814\pm 0.026$ &$ 2.897\pm 0.027$ &$-0.005\pm 0.003$ &SMARTS \\ 
$3102.479$ &$  0.74$ &$ 2.162\pm 0.014$ &$ 2.845\pm 0.024$ &$ 2.795\pm 0.021$ &$ 2.929\pm 0.024$ &$-0.003\pm 0.003$ &SMARTS \\ 
$3105.485$ &$  1.37$ &$ 2.188\pm 0.012$ &$ 2.793\pm 0.018$ &$ 2.786\pm 0.017$ &$ 2.942\pm 0.019$ &$\hm 0.003\pm 0.002$ &SMARTS \\ 
$3237.884$ &$  0.99$ &$ 2.079\pm 0.011$ &$ 2.690\pm 0.018$ &$ 2.707\pm 0.017$ &$ 2.916\pm 0.019$ &$\hm 0.004\pm 0.002$ &SMARTS \\ 
$3240.859$ &$  0.69$ &$ 2.087\pm 0.012$ &$ 2.698\pm 0.019$ &$ 2.684\pm 0.017$ &$ 2.924\pm 0.020$ &$\hm 0.003\pm 0.002$ &SMARTS \\ 
$3245.860$ &$  0.55$ &$ 2.088\pm 0.012$ &$ 2.668\pm 0.018$ &$ 2.676\pm 0.017$ &$ 2.849\pm 0.019$ &$\hm 0.001\pm 0.003$ &SMARTS \\ 
$3248.962$ &$  0.48$ &$ 2.060\pm 0.012$ &$ 2.718\pm 0.019$ &$ 2.665\pm 0.018$ &$ 2.846\pm 0.021$ &$-0.032\pm 0.002$ &  APO \\ 
$3251.810$ &$  0.87$ &$ 2.072\pm 0.013$ &$ 2.676\pm 0.022$ &$ 2.651\pm 0.020$ &$ 2.852\pm 0.023$ &$-0.002\pm 0.003$ &SMARTS \\ 
$3258.833$ &$  1.71$ &$ 2.109\pm 0.010$ &$ 2.646\pm 0.013$ &$ 2.653\pm 0.013$ &$ 2.877\pm 0.015$ &$\hm 0.005\pm 0.002$ &SMARTS \\ 
$3263.793$ &$  2.22$ &$ 2.091\pm 0.009$ &$ 2.688\pm 0.012$ &$ 2.678\pm 0.011$ &$ 2.875\pm 0.013$ &$\hm 0.005\pm 0.002$ &SMARTS \\ 
$3267.836$ &$  0.74$ &$ 2.111\pm 0.012$ &$ 2.656\pm 0.017$ &$ 2.676\pm 0.016$ &$ 2.861\pm 0.018$ &$\hm 0.001\pm 0.003$ &SMARTS \\ 
$3270.816$ &$  1.53$ &$ 2.085\pm 0.009$ &$ 2.661\pm 0.012$ &$ 2.676\pm 0.012$ &$ 2.869\pm 0.014$ &$\hm 0.005\pm 0.002$ &SMARTS \\ 
$3273.787$ &$  1.43$ &$ 2.127\pm 0.012$ &$ 2.659\pm 0.017$ &$ 2.649\pm 0.016$ &$ 2.837\pm 0.019$ &$\hm 0.001\pm 0.003$ &SMARTS \\ 
$3282.705$ &$  1.32$ &$ 2.119\pm 0.011$ &$ 2.652\pm 0.015$ &$ 2.670\pm 0.015$ &$ 2.895\pm 0.018$ &$\hm 0.002\pm 0.002$ &SMARTS \\ 
$3284.804$ &$  0.70$ &$ 2.107\pm 0.013$ &$ 2.701\pm 0.022$ &$ 2.684\pm 0.020$ &$ 2.876\pm 0.023$ &$\hm 0.001\pm 0.003$ &SMARTS \\ 
$3289.712$ &$  1.48$ &$ 2.097\pm 0.012$ &$ 2.691\pm 0.020$ &$ 2.664\pm 0.017$ &$ 2.873\pm 0.020$ &$\hm 0.004\pm 0.002$ &SMARTS \\ 
$3292.776$ &$  1.40$ &$ 2.085\pm 0.012$ &$ 2.737\pm 0.020$ &$ 2.671\pm 0.017$ &$ 2.845\pm 0.020$ &$\hm 0.003\pm 0.003$ &SMARTS \\ 
$3293.773$ &$  0.75$ &$ 2.074\pm 0.011$ &$ 2.687\pm 0.017$ &$ 2.698\pm 0.016$ &$ 2.918\pm 0.018$ &$\hm 0.005\pm 0.002$ &SMARTS \\ 
$3296.777$ &$  1.90$ &$ 2.085\pm 0.009$ &$ 2.705\pm 0.013$ &$ 2.674\pm 0.013$ &$ 2.854\pm 0.014$ &$\hm 0.005\pm 0.002$ &SMARTS \\ 
$3298.703$ &$  0.94$ &$ 2.061\pm 0.011$ &$ 2.666\pm 0.017$ &$ 2.681\pm 0.016$ &$ 2.881\pm 0.019$ &$\hm 0.001\pm 0.003$ &SMARTS \\ 
$3302.659$ &$  0.78$ &$ 2.071\pm 0.019$ &$ 2.640\pm 0.033$ &$ 2.716\pm 0.031$ &$ 2.893\pm 0.036$ &$-0.006\pm 0.003$ &SMARTS \\ 
$3306.684$ &$  0.49$ &$ 2.065\pm 0.015$ &$ 2.617\pm 0.024$ &$ 2.701\pm 0.025$ &$ 2.918\pm 0.029$ &$-0.008\pm 0.003$ &SMARTS \\ 
$3308.711$ &$  0.67$ &$ 2.075\pm 0.013$ &$ 2.674\pm 0.020$ &$ 2.652\pm 0.020$ &$ 2.869\pm 0.023$ &$-0.004\pm 0.003$ &SMARTS \\ 
$3310.673$ &$  1.00$ &$ 2.057\pm 0.010$ &$ 2.645\pm 0.015$ &$ 2.667\pm 0.015$ &$ 2.843\pm 0.017$ &$\hm 0.001\pm 0.002$ &SMARTS \\ 
$3315.652$ &$  1.49$ &$ 2.069\pm 0.009$ &$ 2.650\pm 0.013$ &$ 2.644\pm 0.013$ &$ 2.870\pm 0.015$ &$\hm 0.004\pm 0.002$ &SMARTS \\ 
$3320.696$ &$  1.91$ &$ 2.076\pm 0.009$ &$ 2.652\pm 0.013$ &$ 2.661\pm 0.012$ &$ 2.858\pm 0.014$ &$\hm 0.005\pm 0.002$ &SMARTS \\ 
$3323.637$ &$  0.92$ &$ 2.102\pm 0.011$ &$ 2.658\pm 0.016$ &$ 2.678\pm 0.016$ &$ 2.839\pm 0.017$ &$\hm 0.002\pm 0.003$ &SMARTS \\ 
$3326.603$ &$  0.82$ &$ 2.120\pm 0.012$ &$ 2.645\pm 0.017$ &$ 2.684\pm 0.017$ &$ 2.837\pm 0.018$ &$\hm 0.002\pm 0.002$ &SMARTS \\ 
$3329.682$ &$  1.23$ &$ 2.100\pm 0.009$ &$ 2.656\pm 0.013$ &$ 2.695\pm 0.013$ &$ 2.839\pm 0.014$ &$\hm 0.004\pm 0.002$ &SMARTS \\ 
$3332.594$ &$  0.34$ &$ 2.143\pm 0.028$ &$ 2.626\pm 0.045$ &$ 2.801\pm 0.049$ &$ 2.809\pm 0.048$ &$-0.020\pm 0.003$ &SMARTS \\ 
$3337.677$ &$  1.19$ &$ 2.145\pm 0.017$ &$ 2.713\pm 0.027$ &$ 2.795\pm 0.029$ &$ 2.815\pm 0.029$ &$-0.011\pm 0.003$ &SMARTS \\ 
$3339.622$ &$  0.72$ &$ 2.157\pm 0.014$ &$ 2.703\pm 0.021$ &$ 2.751\pm 0.021$ &$ 2.868\pm 0.023$ &$-0.003\pm 0.003$ &SMARTS \\ 
$3345.555$ &$  1.34$ &$ 2.221\pm 0.012$ &$ 2.744\pm 0.017$ &$ 2.749\pm 0.016$ &$ 2.936\pm 0.018$ &$\hm 0.004\pm 0.002$ &SMARTS \\ 
$3348.651$ &$  1.31$ &$ 2.222\pm 0.010$ &$ 2.745\pm 0.013$ &$ 2.786\pm 0.013$ &$ 2.926\pm 0.015$ &$\hm 0.005\pm 0.002$ &SMARTS \\ 
$3349.714$ &$  0.69$ &$ 2.212\pm 0.014$ &$ 2.741\pm 0.022$ &$ 2.802\pm 0.020$ &$ 2.956\pm 0.023$ &$-0.032\pm 0.002$ &  APO \\ 
$3351.639$ &$  0.75$ &$ 2.242\pm 0.011$ &$ 2.759\pm 0.015$ &$ 2.810\pm 0.015$ &$ 2.957\pm 0.017$ &$\hm 0.003\pm 0.002$ &SMARTS \\ 
$3353.584$ &$  1.54$ &$ 2.246\pm 0.010$ &$ 2.756\pm 0.013$ &$ 2.802\pm 0.013$ &$ 2.967\pm 0.014$ &$\hm 0.005\pm 0.002$ &SMARTS \\ 
$3354.670$ &$  1.50$ &$ 2.239\pm 0.010$ &$ 2.798\pm 0.014$ &$ 2.815\pm 0.014$ &$ 2.961\pm 0.015$ &$\hm 0.005\pm 0.002$ &SMARTS \\ 
$3355.660$ &$  2.50$ &$ 2.242\pm 0.009$ &$ 2.808\pm 0.013$ &$ 2.838\pm 0.012$ &$ 2.965\pm 0.013$ &$\hm 0.006\pm 0.002$ &SMARTS \\ 
$3356.647$ &$  1.71$ &$ 2.229\pm 0.010$ &$ 2.818\pm 0.015$ &$ 2.829\pm 0.014$ &$ 2.956\pm 0.015$ &$\hm 0.005\pm 0.002$ &SMARTS \\ 
$3357.582$ &$  1.48$ &$ 2.258\pm 0.011$ &$ 2.788\pm 0.015$ &$ 2.804\pm 0.014$ &$ 2.982\pm 0.016$ &$\hm 0.003\pm 0.002$ &SMARTS \\ 
$3357.708$ &$  0.54$ &$ 2.232\pm 0.011$ &$ 2.806\pm 0.017$ &$ 2.812\pm 0.015$ &$ 2.982\pm 0.017$ &$-0.027\pm 0.002$ &  APO \\ 
$3358.630$ &$  1.71$ &$ 2.233\pm 0.010$ &$ 2.792\pm 0.014$ &$ 2.803\pm 0.013$ &$ 2.994\pm 0.015$ &$\hm 0.004\pm 0.002$ &SMARTS \\ 
$3359.624$ &$  1.75$ &$ 2.219\pm 0.010$ &$ 2.784\pm 0.015$ &$ 2.798\pm 0.014$ &$ 2.954\pm 0.016$ &$\hm 0.003\pm 0.002$ &SMARTS \\ 
$3360.595$ &$  0.96$ &$ 2.218\pm 0.014$ &$ 2.794\pm 0.021$ &$( 2.750\pm 0.020)$ &$ 3.020\pm 0.024$ &$-0.001\pm 0.003$ &SMARTS \\ 
$3361.625$ &$  0.82$ &$ 2.206\pm 0.011$ &$( 2.767\pm 0.016)$ &$ 2.775\pm 0.016$ &$ 2.978\pm 0.018$ &$\hm 0.002\pm 0.002$ &SMARTS \\ 
$3362.605$ &$  0.63$ &$ 2.212\pm 0.013$ &$ 2.794\pm 0.020$ &$ 2.797\pm 0.020$ &$ 2.996\pm 0.023$ &$-0.001\pm 0.003$ &SMARTS \\ 
$3366.603$ &$  1.16$ &$ 2.203\pm 0.017$ &$ 2.781\pm 0.027$ &$ 2.779\pm 0.027$ &$ 3.044\pm 0.033$ &$-0.008\pm 0.003$ &SMARTS \\ 
$3367.669$ &$  1.61$ &$ 2.178\pm 0.013$ &$ 2.846\pm 0.022$ &$ 2.841\pm 0.021$ &$ 2.956\pm 0.023$ &$-0.003\pm 0.003$ &SMARTS \\ 
$3369.662$ &$  0.93$ &$ 2.197\pm 0.011$ &$ 2.831\pm 0.018$ &$ 2.788\pm 0.017$ &$ 3.022\pm 0.020$ &$\hm 0.001\pm 0.002$ &SMARTS \\ 
$3370.637$ &$  1.44$ &$ 2.204\pm 0.011$ &$ 2.857\pm 0.016$ &$ 2.815\pm 0.015$ &$ 3.011\pm 0.017$ &$\hm 0.004\pm 0.002$ &SMARTS \\ 
$3371.592$ &$  1.64$ &$ 2.208\pm 0.010$ &$ 2.839\pm 0.015$ &$ 2.772\pm 0.014$ &$ 3.026\pm 0.016$ &$\hm 0.005\pm 0.002$ &SMARTS \\ 
$3372.609$ &$  1.07$ &$ 2.179\pm 0.010$ &$ 2.819\pm 0.015$ &$ 2.801\pm 0.014$ &$ 3.001\pm 0.016$ &$\hm 0.003\pm 0.002$ &SMARTS \\ 
$3373.634$ &$  0.92$ &$ 2.180\pm 0.012$ &$ 2.823\pm 0.020$ &$ 2.797\pm 0.018$ &$ 3.036\pm 0.021$ &$\hm 0.003\pm 0.002$ &SMARTS \\ 
$3374.591$ &$  1.06$ &$ 2.167\pm 0.012$ &$ 2.861\pm 0.021$ &$ 2.800\pm 0.018$ &$ 3.036\pm 0.022$ &$\hm 0.002\pm 0.003$ &SMARTS \\ 
$3375.586$ &$  2.01$ &$ 2.153\pm 0.010$ &$ 2.834\pm 0.015$ &$ 2.754\pm 0.013$ &$ 3.012\pm 0.016$ &$\hm 0.005\pm 0.002$ &SMARTS \\ 
$3376.612$ &$  1.27$ &$ 2.145\pm 0.009$ &$ 2.799\pm 0.014$ &$ 2.766\pm 0.013$ &$ 3.041\pm 0.015$ &$\hm 0.005\pm 0.002$ &SMARTS \\ 
$3377.603$ &$  1.58$ &$ 2.145\pm 0.010$ &$ 2.815\pm 0.014$ &$ 2.773\pm 0.013$ &$ 3.030\pm 0.016$ &$\hm 0.005\pm 0.002$ &SMARTS \\ 
$3378.636$ &$  1.03$ &$ 2.127\pm 0.010$ &$ 2.837\pm 0.015$ &$ 2.752\pm 0.014$ &$ 3.032\pm 0.016$ &$\hm 0.004\pm 0.002$ &SMARTS \\ 
$3379.620$ &$  0.90$ &$ 2.149\pm 0.010$ &$ 2.769\pm 0.016$ &$ 2.729\pm 0.014$ &$ 3.028\pm 0.017$ &$\hm 0.004\pm 0.002$ &SMARTS \\ 
$3380.602$ &$  1.50$ &$ 2.133\pm 0.010$ &$ 2.780\pm 0.015$ &$ 2.714\pm 0.013$ &$ 3.041\pm 0.017$ &$\hm 0.004\pm 0.002$ &SMARTS \\ 
$3381.588$ &$  2.06$ &$ 2.128\pm 0.009$ &$ 2.801\pm 0.013$ &$ 2.703\pm 0.012$ &$ 3.033\pm 0.015$ &$\hm 0.005\pm 0.002$ &SMARTS \\ 
$3382.583$ &$  1.01$ &$ 2.103\pm 0.011$ &$ 2.756\pm 0.016$ &$ 2.697\pm 0.015$ &$ 3.031\pm 0.019$ &$\hm 0.002\pm 0.003$ &SMARTS \\ 
$3383.611$ &$  2.31$ &$ 2.098\pm 0.009$ &$ 2.741\pm 0.012$ &$ 2.728\pm 0.012$ &$ 3.027\pm 0.014$ &$\hm 0.005\pm 0.002$ &SMARTS \\ 
$3385.591$ &$  0.94$ &$ 2.081\pm 0.012$ &$ 2.759\pm 0.019$ &$ 2.752\pm 0.018$ &$ 3.010\pm 0.021$ &$\hm 0.000\pm 0.003$ &SMARTS \\ 
$3386.601$ &$  1.92$ &$ 2.064\pm 0.009$ &$ 2.774\pm 0.014$ &$ 2.748\pm 0.013$ &$ 3.009\pm 0.016$ &$\hm 0.004\pm 0.002$ &SMARTS \\ 
$3387.592$ &$  1.37$ &$ 2.081\pm 0.010$ &$ 2.750\pm 0.014$ &$ 2.695\pm 0.014$ &$ 2.970\pm 0.016$ &$\hm 0.003\pm 0.002$ &SMARTS \\ 
$3394.604$ &$  0.58$ &$ 2.077\pm 0.014$ &$ 2.717\pm 0.022$ &$ 2.668\pm 0.021$ &$ 2.934\pm 0.026$ &$-0.005\pm 0.003$ &SMARTS \\ 
$3395.603$ &$  0.90$ &$ 2.068\pm 0.013$ &$ 2.746\pm 0.022$ &$ 2.681\pm 0.020$ &$ 2.868\pm 0.023$ &$-0.004\pm 0.003$ &SMARTS \\ 
$3396.540$ &$  0.72$ &$ 2.064\pm 0.015$ &$ 2.712\pm 0.025$ &$ 2.719\pm 0.024$ &$ 2.907\pm 0.027$ &$-0.006\pm 0.003$ &SMARTS \\ 
$3397.559$ &$  0.86$ &$ 2.053\pm 0.010$ &$ 2.711\pm 0.016$ &$ 2.673\pm 0.015$ &$ 2.914\pm 0.018$ &$\hm 0.001\pm 0.002$ &SMARTS \\ 
$3398.553$ &$  2.26$ &$ 2.034\pm 0.009$ &$ 2.748\pm 0.013$ &$ 2.695\pm 0.012$ &$ 2.878\pm 0.014$ &$\hm 0.004\pm 0.002$ &SMARTS \\ 
$3399.577$ &$  2.90$ &$ 2.040\pm 0.009$ &$ 2.732\pm 0.014$ &$ 2.717\pm 0.013$ &$ 2.861\pm 0.014$ &$\hm 0.004\pm 0.002$ &SMARTS \\ 
$3402.527$ &$  0.93$ &$ 2.039\pm 0.011$ &$ 2.744\pm 0.019$ &$ 2.686\pm 0.018$ &$ 2.866\pm 0.020$ &$-0.001\pm 0.003$ &SMARTS \\ 
$3403.559$ &$  1.40$ &$ 2.037\pm 0.009$ &$ 2.690\pm 0.013$ &$ 2.678\pm 0.013$ &$ 2.873\pm 0.014$ &$\hm 0.004\pm 0.002$ &SMARTS \\ 
$3404.573$ &$  1.13$ &$ 2.043\pm 0.011$ &$ 2.683\pm 0.016$ &$ 2.722\pm 0.015$ &$ 2.891\pm 0.017$ &$\hm 0.002\pm 0.002$ &SMARTS \\ 
$3405.560$ &$  1.60$ &$ 2.051\pm 0.010$ &$ 2.692\pm 0.014$ &$ 2.679\pm 0.014$ &$ 2.893\pm 0.016$ &$\hm 0.003\pm 0.002$ &SMARTS \\ 
$3406.563$ &$  1.68$ &$ 2.051\pm 0.010$ &$ 2.707\pm 0.014$ &$ 2.700\pm 0.014$ &$ 2.861\pm 0.015$ &$\hm 0.004\pm 0.002$ &SMARTS \\ 
$3407.560$ &$  0.82$ &$ 2.064\pm 0.011$ &$ 2.691\pm 0.016$ &$ 2.695\pm 0.015$ &$ 2.898\pm 0.017$ &$\hm 0.003\pm 0.003$ &SMARTS \\ 
$3408.573$ &$  1.31$ &$ 2.057\pm 0.010$ &$ 2.684\pm 0.014$ &$ 2.679\pm 0.014$ &$ 2.871\pm 0.015$ &$\hm 0.004\pm 0.002$ &SMARTS \\ 
$3409.559$ &$  1.02$ &$ 2.060\pm 0.010$ &$ 2.705\pm 0.015$ &$ 2.683\pm 0.014$ &$ 2.887\pm 0.016$ &$\hm 0.003\pm 0.002$ &SMARTS \\ 
$3410.557$ &$  2.02$ &$ 2.059\pm 0.009$ &$ 2.662\pm 0.012$ &$ 2.678\pm 0.012$ &$ 2.887\pm 0.013$ &$\hm 0.005\pm 0.002$ &SMARTS \\ 
$3411.560$ &$  1.04$ &$ 2.057\pm 0.010$ &$ 2.683\pm 0.015$ &$ 2.716\pm 0.015$ &$ 2.885\pm 0.016$ &$\hm 0.004\pm 0.002$ &SMARTS \\ 
$3413.559$ &$  1.40$ &$ 2.058\pm 0.011$ &$ 2.704\pm 0.018$ &$ 2.720\pm 0.017$ &$ 2.839\pm 0.018$ &$\hm 0.002\pm 0.002$ &SMARTS \\ 
$3414.548$ &$  1.04$ &$ 2.067\pm 0.013$ &$ 2.714\pm 0.021$ &$ 2.674\pm 0.018$ &$ 2.820\pm 0.020$ &$-0.001\pm 0.003$ &SMARTS \\ 
$3415.546$ &$  2.39$ &$ 2.060\pm 0.010$ &$ 2.695\pm 0.014$ &$ 2.711\pm 0.013$ &$ 2.837\pm 0.015$ &$\hm 0.003\pm 0.002$ &SMARTS \\ 
$3420.566$ &$  0.76$ &$ 2.065\pm 0.013$ &$ 2.690\pm 0.020$ &$ 2.673\pm 0.019$ &$ 2.888\pm 0.022$ &$-0.002\pm 0.003$ &SMARTS \\ 
$3422.590$ &$  0.67$ &$ 2.041\pm 0.014$ &$ 2.677\pm 0.024$ &$ 2.682\pm 0.023$ &$ 2.910\pm 0.028$ &$-0.006\pm 0.003$ &SMARTS \\ 
$3426.545$ &$  0.85$ &$ 2.058\pm 0.017$ &$ 2.736\pm 0.031$ &$ 2.749\pm 0.028$ &$ 2.846\pm 0.031$ &$-0.006\pm 0.003$ &SMARTS \\ 
$3427.548$ &$  0.51$ &$ 2.074\pm 0.013$ &$ 2.676\pm 0.022$ &$ 2.691\pm 0.021$ &$ 2.897\pm 0.024$ &$-0.003\pm 0.003$ &SMARTS \\ 
$3432.535$ &$  2.14$ &$ 2.049\pm 0.010$ &$ 2.728\pm 0.016$ &$ 2.733\pm 0.015$ &$ 2.865\pm 0.016$ &$\hm 0.003\pm 0.002$ &SMARTS \\ 
$3433.538$ &$  2.87$ &$ 2.044\pm 0.011$ &$ 2.689\pm 0.017$ &$ 2.711\pm 0.016$ &$ 2.827\pm 0.017$ &$\hm 0.002\pm 0.002$ &SMARTS \\ 
$3435.535$ &$  1.05$ &$ 2.033\pm 0.011$ &$ 2.686\pm 0.017$ &$ 2.700\pm 0.016$ &$ 2.853\pm 0.018$ &$\hm 0.002\pm 0.003$ &SMARTS \\ 
$3439.584$ &$  0.40$ &$ 2.022\pm 0.012$ &$ 2.668\pm 0.020$ &$( 2.626\pm 0.018)$ &$ 2.852\pm 0.022$ &$-0.034\pm 0.002$ &  APO \\ 
$3444.512$ &$  0.83$ &$ 2.010\pm 0.013$ &$ 2.739\pm 0.022$ &$ 2.681\pm 0.020$ &$ 2.838\pm 0.022$ &$-0.002\pm 0.003$ &SMARTS \\ 
$3446.492$ &$  1.49$ &$ 1.935\pm 0.019$ &$ 2.729\pm 0.040$ &$ 2.782\pm 0.038$ &$ 2.845\pm 0.040$ &$-0.014\pm 0.003$ &SMARTS \\ 
$3448.503$ &$  1.46$ &$ 1.991\pm 0.011$ &$ 2.699\pm 0.019$ &$ 2.658\pm 0.017$ &$ 2.847\pm 0.020$ &$-0.001\pm 0.003$ &SMARTS \\ 
$3453.506$ &$  0.68$ &$ 1.995\pm 0.014$ &$ 2.647\pm 0.025$ &$ 2.688\pm 0.024$ &$ 2.852\pm 0.027$ &$-0.007\pm 0.003$ &SMARTS \\ 
$3460.480$ &$  0.68$ &$ 1.934\pm 0.023$ &$ 2.694\pm 0.048$ &$ 2.736\pm 0.042$ &$ 2.849\pm 0.047$ &$-0.015\pm 0.003$ &SMARTS \\ 
$3464.483$ &$  0.79$ &$ 1.936\pm 0.016$ &$ 2.661\pm 0.032$ &$ 2.739\pm 0.029$ &$ 2.816\pm 0.031$ &$-0.006\pm 0.003$ &SMARTS \\ 
$3469.479$ &$  1.46$ &$ 1.940\pm 0.015$ &$ 2.745\pm 0.029$ &$ 2.746\pm 0.025$ &$ 2.768\pm 0.025$ &$-0.005\pm 0.003$ &SMARTS \\ 

\enddata
\tablecomments{The HJD is relative to JD 2450000.
  The formal uncertainties were rescaled whenever the value
  of $\chi^2/N_{dof}$ was greater than unity (see text).  The QSO A-D
  columns give the magnitudes of the quasar images relative to the
  comparison stars. The $\langle\hbox{Stars}\rangle$ column gives the
  mean magnitude of the standard stars for that epoch relative to
  their mean for all epochs.  The mean is non-zero because of the 
  structure of the covariance matrix, but it should deviate from zero
  by no more than its uncertainties.  Note that
  there is a small offset of the APO r-band fluxes from the SMARTS
  R-band fluxes (see text). A few points, enclosed in parentheses, are
  dropped in the fits to the light curves.  }
\label{tab:lightcurves}
\end{deluxetable}

\begin{deluxetable}{ccccc}
\tablecaption{Microlensing parameters for \he0435}
\tablewidth{0pt}
\tablehead{
    \multicolumn{1}{c}{Image}
   &\multicolumn{1}{c}{Season}
   &\multicolumn{1}{c}{$\Delta m$}
   &\multicolumn{1}{c}{$d\Delta m/dt$}
   &\multicolumn{1}{c}{$d^2\Delta m/dt^2$} \\
    \multicolumn{1}{c}{Pair}
   &
   &\multicolumn{1}{c}{(mag)}
   &\multicolumn{1}{c}{(mag/yr)}
   &\multicolumn{1}{c}{(mag/yr$^2$)} 
                 }
\startdata
AB  & 1 &$\hm0.584\pm0.004$ &$\hm0.161\pm0.021$ &$\hm0.465\pm0.212$  \\ 
    & 2 &$\hm0.623\pm0.004$ &$\hm0.206\pm0.017$ &$\hm1.226\pm0.185$  \\ 
AC  & 1 &$\hm0.571\pm0.004$ &$\hm0.199\pm0.020$ &$-0.051\pm0.195$  \\ 
    & 2 &$\hm0.624\pm0.003$ &$\hm0.187\pm0.016$ &$\hm1.473\pm0.170$  \\ 
AD  & 1 &$\hm0.740\pm0.004$ &$\hm0.037\pm0.021$ &$-0.192\pm0.220$  \\ 
    & 2 &$\hm0.795\pm0.004$ &$\hm0.082\pm0.019$ &$\hm0.060\pm0.202$  \\ 

\enddata
\tablecomments{The parameters of the model of differential
microlensing variability in \he0435.  The uncertainties do not include
uncertainties arising from the uncertainties in the time delays.  }
\label{tab:micro}
\end{deluxetable}

\end{document}